\documentclass[aps,pra,twocolumn,preprintnumbers,amssymb,amsfonts,amsmath,superscriptaddress,showpacs,hyperref]{revtex4-1}

\pdfoutput=1
\usepackage{amssymb}
\usepackage{cancel}
\usepackage{color,graphicx}
\usepackage{amsmath}
\usepackage{amsbsy}
\usepackage{amsthm}
\usepackage{bbm}
\usepackage{epsfig}
\usepackage{float}
\usepackage{graphicx}
\usepackage{dcolumn}
\usepackage{bbm}
\usepackage{color,epstopdf}
\usepackage{amscd}
\usepackage{amsfonts}  
\usepackage[]{amsmath}
\usepackage{amssymb}    
\usepackage{mathrsfs}
\usepackage{verbatim}
\usepackage[]{cases}
\usepackage{amsmath}
\usepackage{wasysym}
\usepackage[utf8]{inputenc}
\usepackage[T1]{fontenc}
\usepackage{mathtools}
\usepackage{xcolor}
\usepackage{tabularx}

\newcommand{\hide}[1]{\relax}

\newcommand{\nocontentsline}[3]{}
\newcommand{\tocless}[2]{\bgroup\let\addcontentsline=\nocontentsline#1{#2}\egroup}

\newcommand{\I}{\text{\tiny{I}}}

\DeclareMathOperator{\sign}{sign}

\begin{document}

\title{Prospects for near-field interferometric tests of Collapse Models}

\author{Giulio Gasbarri}\thanks{These authors contributed equally to this work}
\affiliation{School of Physics and Astronomy, University of Southampton, SO17 1BJ, United Kingdom}

\affiliation{F\'isica Te\`orica: Informaci\'o i Fen\`omens Qu\`antics, Department de F\'isica, Universitat Aut\`onoma de Barcelona, 08193 Bellaterra (Barcelona), Spain}
\email{Giulio.Gasbarri@uab.cat}
\author{Alessio Belenchia}
\thanks{These authors contributed equally to this work}
\affiliation{Centre for Theoretical Atomic, Molecular, and Optical Physics, School of Mathematics and Physics, Queens University, Belfast BT7 1NN, United Kingdom}
\email{A.Belenchia@qub.ac.uk}
\author{Mauro Paternostro}
\affiliation{Centre for Theoretical Atomic, Molecular, and Optical Physics, School of Mathematics and Physics, Queens University, Belfast BT7 1NN, United Kingdom}
\author{Hendrik Ulbricht}
\email{H.Ulbricht@soton.ac.uk}
\affiliation{School of Physics and Astronomy, University of Southampton, SO17 1BJ, United Kingdom}

\begin{abstract}
{Near-field interferometry with large dielectric nano-particles opens the way to test fundamental modification of standard quantum mechanics at an unprecedented level. We showcase the capabilities of such platform, in a state-of-the-art ground-based experimental set-up, to set new stringent bounds on the parameters space of collapse models and highlight the future perspective for this class of experiments.}
\end{abstract}
\maketitle

\section{Introduction}


 
The problem of how the deterministic classical world that we experience emerges from a probabilistic microcopic quantum world has puzzled physicists since the early days of quantum mechanics. Several ideas have been proposed to solve this conundrum, ranging from the superposition of macroscopic systems being hard to produce and maintain due to the unavoidable interaction with the surrounding environment~\cite{joos1985emergence,caldeira1983path,zurek1982environment,zurek1981pointer,zeh1973toward,zeh1970interpretation}, to there being a more fundamental mechanism which prevents the macroscopic world from behaving quantum mechanically.
 
One of the most significant of such proposals invokes the so-called spontaneous localization models, also known 
as collapse models (CM)~\cite{bassi2003dynamical,ghirardi1990markov,ghirardi1986unified}.
In the collapse-model framework, the unitary Schr\"odinger dynamics of close quantum systems has to be modified by a stochastic collapse of the wave-function. Such mechanism should be such that the microscopic predictions of quantum mechanics are preserved, while macroscopic objects abide classical physics. By promoting the collapse to a physical process, and removing the need to introduce \textit{ad hoc} postulates for measurement events, collapse models \textit{de facto} resolve the measurement problem of quantum mechanics. 
However, they are phenomenological models and present free parameters whose values should be constrained by experiments. In order to do so, one needs to look at the 
net effect of the stochastic collapse of the wave-function postulated by CM, 
which is an extra source of decoherence beyond the environmental ones.

As CMs imply that the quantum mechanical superposition principle is not universally valid, matter-wave interference experiments represent excellent candidates for testing these ideas. On the one hand, they provide a direct test of the quantum superposition principle. On the other hand, their sensitivity allows to set bounds to any alleged modification of quantum theory, including  mechanisms such as CMs.
%
This is at variance with non-interferometric schemes (see~~\cite{carlesso2019collapse} and references therein) which have been the power-horse for test of CMs to date but cannot provide a direct test of the quantum superposition principle.


Recently, Ref.~\cite{fein2019quantum} reported an  experimental scheme, based on a Talbot-Lau interferometer~\cite{gerlich2007kapitza,brezger2003concepts}, testing the superposition principle on a 25 amu (atomic mass unit) mass and, at the same time, providing new bounds on the mechanism for quantum-to-classical transition entailed by CMs. 
Recent progress in near-field interferometry with optical grating and in the trapping and cooling of nanoparticles in optical cavities, have led to realistic proposals for ground- ~\cite{bateman2014near} (and space-based~\cite{kaltenbaek2016macroscopic})  experiments pushing the masses of the nanoparticles used for such experiments up to $10^{6}$amu ($10^{11}$amu in space), with the perspective of exploring quantum superpositions of objects of relatively large size and mass.

These advancements have placed interferometric experiments at the frontier of experimental tests of CMs and elevated matter-wave interferometry as a very promising platform to directly test the superposition principle at increasing scales. 

In this paper, we aim to study the potential of table-top near-field interferometry to 
constrain the parameter space of the dynamics set by CMs, and with that test the superposition principle of quantum mechanics.
We {consider single-grating matter-wave Talbot interferometry set-ups~\cite{bateman2014near}}
for a ground-based experiment with nanoparicles with mass of $10^6$~amu {, which are well within current technological possibilities,} and show the possible stringent bounds on the CM parameter space that can be achieved in such a way. 
We, furthermore, study the case of nanoparticles with mass up to $10^{7}$~amu, outside of the Rayleigh approximation, employing the methods developed in~\cite{belenchia2019talbot}.

We perform our analysis by including all known sources of environmental decoherence and spanning the relevant parameter space of the continuous spontaneous localization (CSL) model, which is one of the currently most studied CM~\cite{carlesso2019collapse}. In this way, we are able to provide the first ab-initio estimate of the possibilities offered by near-field interferometry ground-based experiments, employing large dielectric nanoparticles, for testing deviations from quantum mechanics. In particular, we show that there is a window of opportunity to challenge CMs up to the historical GRW parameter values. We also present an in-depth discussion of current technological challenges, and possible solutions, in this endeavour.

The remainder of this paper is organized as follows. In Sec.~\ref{scheme} we describe the near-field interferometric setting that we analyze. Sec.~\ref{test} provides an analytic study of the possibilities offered by such setting to test the CSL model. In Sec.~\ref{perspectives} we provide a careful assessment fo the experimental route towards the implementation of such test. Finally, in Sec.~\ref{conc} we draw our conclusions. 

\section{Schematic near-field interferometer set-up}
\label{scheme}
\begin{figure}[t!]
\centering
\includegraphics[width=0.85\linewidth]{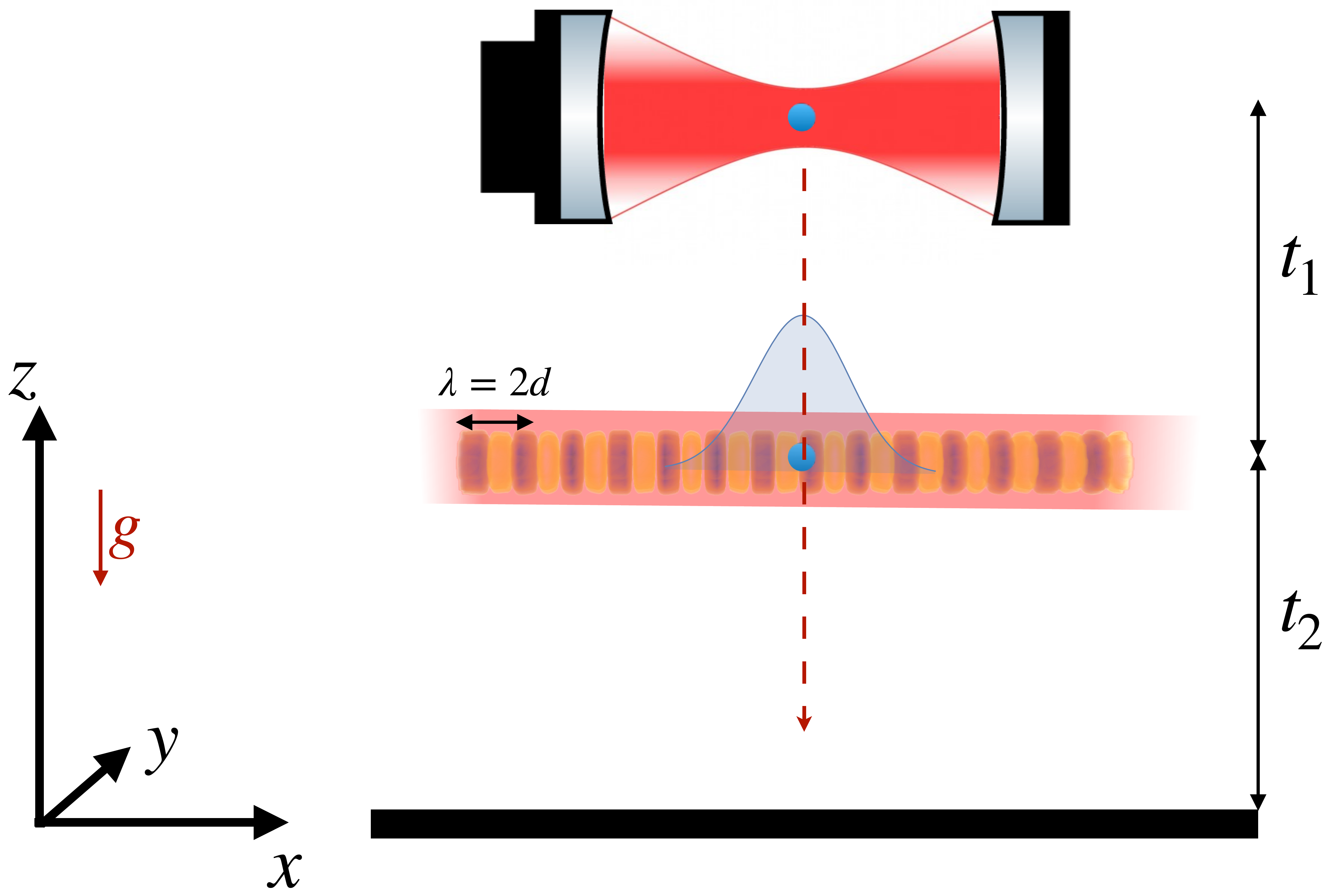}
\caption{Setup of the near-field interferometer with a levitated nanoparticle. The latter is initially trapped and cooled down in an optical cavity. After the cavity is switched off, the particle freely falls for a time $t_1$ before reaching an optical grating generated by a pulsed UV laser that is retro-reflected by a mirror. The grating period is half the wavelength of the standing wave. At the optical grating, the particle is required to have a coherence length able to cover at least two adjacent ``slits'' of the grating. After the grating, the particle freely falls for a time $t_2$ before being measured on the final screen.}
\label{setup}
\end{figure}

{Let us consider the interferometric set-up firstly proposed in Ref.~\cite{bateman2014near} a sketch of which is presented in Fig.~\ref{setup}. At variance with interferometric experiments with lighter particles, where particles beams are engineered, each nanoparticle in the experiment is individually addressed. In each run of the \textit{proposed experiment}: (i) The nanoparticle is firstly trapped and cooled down in an optical cavity, which prepares the particle in a thermal state of its center of mass degree of freedom at $\sim20$mK; (ii) After the cooling period ($t_c$), the cavity is turned off and the particle freely falls for a time $t_1$ before reaching an optical grating; (iii) The  grating, provided by a standing wave of wavelength $\lambda$, is formed by a retro-reflected pulsed UV laser. The matter-light interaction at the optical grating has been shown to be equivalent to a pure phase-grating~\cite{nimmrichter2014macroscopic}. (iv) After the grating, the particle freely evolves again for a time $t_2$ before getting captured on a screen, which records its position.}

{
By repeating this protocol several times, an interference pattern can be formed on the screen. Note that, even by neglecting all decoherence effects, the building up of an interference pattern is possible only if two conditions are satisfied: 
\begin{enumerate}
\item[{\bf (a)}] The post-cooling state of the particle is such that free evolution for a time $t_1$ results in a coherence-length sufficient to cover at least two adjacent "slits" of the optical grating.
\item[{\bf (b)}] The free evolution time $t_2$ is long enough for interference to take place. 
\end{enumerate}
Furthermore, an additional complication arises in near-field interferometry. Indeed, a sample of particles following  ballistic trajectories would form, in a near-field interferometer such as the one described above, a seemingly interferometric pattern due to classical deflection at the optical grating. It is thus crucial to work in a regime in which the predicted quantum interference figure, considering all decoherence sources, is clearly distinguishable from the classical shadow pattern.}

{
The specifics reported in Table~\ref{tab:parameters}, and used throughout this work, guarantee both meeting conditions {\bf (a)} and {\bf (b)}, and the distinguishability of the quantum mechanical pattern from the classical shadow one (see also Fig.~\ref{fig:SiSiO2}), for silicon (Si) and glass silica (SiO2) spherical nanoparticles with a mass of  $10^6$~amu and $10^7$~amu. Note that, while challenging, these working points are within reach of current technology~\cite{bateman2014near}. 
}

{
Finally, it should be noted that, for the case of nanoparticles with masses around $10^6$ amu, the wavelength of the grating laser in Table~\ref{tab:parameters} is such that $2\pi R/\lambda\ll 1$. This allows to treat the particle-light interaction in the Rayleigh limit~\cite{nimmrichter2014macroscopic}. Conversely, the same is not true for particle with mass $10^7$~amu for which we need to apply the formalism recently developed in Ref.~\cite{belenchia2019talbot} to correctly account for the grating interaction.}



\begin{table}[h]
\centering
\begin{tabular}{llll}
    \hline
    \textbf{Symbol}  & & \textbf{Name} & \textbf{Value}  \\ 
    &&&{\bf or Expression}\\
    \hline\hline
    Nanosphere\\ properties:\\ \hline
    $\rho_{Si}$ & & Si density & $2329$~kg/m$^3$ \\
    $\rho_{SiO2}$ & & Glass SiO2 density & $1850$~kg/m$^3$ \\
    $c_{m}$ & & Specific heat  & 700~J/(kg K) \\
    $I$ & & Ionization energy & $5\times 10^{-19}$~J \\
    $m$ & & Mass & $10^6 - 10^7$~amu \\ 
    $\sigma_x$ & & Position variance & $\sqrt{\frac{\hbar}{4\gamma}\coth{({\beta_0\nu_m})}}$ \\
    &&post-cooling &\\
    $\sigma_p$ & & Momentum variance& $\sqrt{\hbar\gamma\coth{({\beta_0\nu_m})}}$ \\
    &&post-cooling &\\
    \hline 
    Trapping\\ parameters:  \\
    \hline
    $\lambda_c$ & & Trap's laser wavelength & 1550~nm \\
    $t_c$ & & Trapping and cooling time & 1~s \\
    $I_{{\rm trap}}$ & & Trap's laser intensity & $90\times 10^9$~W/m$^2$ \\
    $\nu_{m}$ & & Trap mechanical frequency & $200$~Hz \\
    $T_{{\rm int}0}$ & & C.o.m. temperature  & $20$~mK \\
    &&post-cooling &\\
    \hline
    Optical grating\\ parameters: \\ \hline
    $\lambda$ & & Grating laser wavelength & $355$~nm \\
    $E_L/a_L$ & & Pulse-energy per spot-area & cf. Eqs.~(20)-(22)  \\ 
    &&&in Ref.~\cite{belenchia2019talbot}\\
    \hline
    Environment\\ parameters: \\ \hline
    $T_{\rm env}$ & & Environmental temperature & 4 --300~K \\
    $\alpha_g/(4\pi\epsilon_0)$ & &  Residual gas polarizability & $1.74\times 10^{-30}$~m$^3$ \\
    $\I_g$ & & Residual gas ioniz. energy & $15.6\times 10^{-19}$~J\\
    $m_g$ & & Residual gas mass & 28~amu\\
    $v_g$ & & Residual gas mean velocity  & $\sqrt{2k_B T_{\rm env}/m_g}$\\
    $P_g$ & & Residual gas pressure & $10^{-10}$~mbar \\ \hline
\end{tabular}
\caption{Specifics of all the parameters entering the simulation of the near-field interferometric experiments with dielectric nanospheres. The residual gas in the vacuum chamber is assumed to be composed mostly by nitrogen, of which we use the physical properties. The refractive index, as a function of the frequency, for Si and SiO2 can be found tabulated in the supplementary material of~\cite{bateman2014near} (see also references therein). The pulse energy to spot area ratio of the grating laser is related to the eikonal phase by Eqs.~(20-22) in~\cite{belenchia2019talbot} to which we refer the reader for additional details. 
We have set $\beta_0=h/(2k_B T_{{\rm int}0})$ and $\gamma=\pi m\nu_m$.}
\label{tab:parameters}
\end{table}

\section{Tests of CSL Collapse Models}
\label{test}
Collapse models introduce a stochastic collapse of the wave function into the Schr\"odinger equation driven by a fundamental noise field, usually assumed to be white-noise (see~\cite{gasbarri2017gravity,ferialdi2012dissipative,adler2008collapse,adler2007collapse} for generalization to colored-noises).
However, the stochastic Schr$\ddot{\text{o}}$dinger equation is not directly addressable and the net result of the CM is the presence of a decoherent dynamics of the closed quantum system, that, in the case of study, is described by the following center of mass master equation~\cite{bassi2003dynamical}
\begin{equation}\label{cslme}
\partial_{t}{\hat{\rho}}_{t}^{\text{\tiny CM}}=-\frac{i}{\hbar}\left[\hat{H}_{0}^{\textrm{\tiny CM}},\hat{\rho}_{t}^{\text{\tiny CM}}\right]-{\cal D}(\hat\rho^{\text{CM}}_t)
\end{equation}
where we have introduced the CM-induced non-unitary term
\begin{equation}
\label{decoCM}
{\cal D}(\hat\rho^{\rm CM}_t)=\frac{\lambda(4\pi r_c^2)^{\frac{3}{2}}}{(2\pi\hbar)^{3}m^2_0}\!\!\int\!\! d\mathbf{q}\,{\tilde{\mu}(\mathbf{q})}\,e^{-\frac{r_c^2\mathbf{q}^{2}}{\hbar^{2}}}\,\hspace{-0.15cm}\left[e^{-\frac{i}{\hbar}\mathbf{q}\cdot\hat{\mathbf{x}}},\!\left[e^{\frac{i}{\hbar}\mathbf{q}\cdot\hat{\mathbf{x}}}\!,\hat{\rho}_{t}^{\textrm{\tiny CM}}\right]\right]
    \end{equation}
with $m_{0}$ the nucleon mass, $\lambda$ and $r_{c}$ the rate and localization distance of the CM respectively, and  
\begin{align}
\tilde{\mu}(\mathbf{q}):= \int d \mathbf{x}\, e^{-\frac{i}{\hbar}\mathbf{q}\cdot\mathbf{x}}\mu(\mathbf{x})    
\end{align}
the Fourier transform of the particle mass distribution $\mu(\mathbf{x})$. Both $r_C$ and $\lambda$ are free parameters of the CM. According to Eq.~\eqref{decoCM}, decoherence occurs in the position basis and it is stronger the more massive the system becomes, giving rise to spatial localization of macroscopic quantum systems and thus suppressing macroscopic spatial superposition~\cite{bassi2003dynamical}.   

In the case of interest, the nanoparticles in the interferometric experiments are not completely isolated from the external environment. Indeed, they are subject to several decoherence channels in addition to, and in competition with, the alleged CM noise. In practice, the experiment can ``only'' aim at casting upper bounds on the parameters of the CM by looking at excess decoherence with respect to the predicted one from the environment. While any excess decoherence, with respect to the theoretical prediction, cannot be interpreted straight away as evidence of the occurrence of CM, we would like to emphasize that a consistent observation of such extra decoherence with different experimental parameters --- or in several different experimental platforms --- and in agreement with the predicted scaling with the parameters of the system, could be claimed to be a righteous verification of CM theory. 

In our simulation of the near-field interferometer, we have modelled several sources of decoherence, including mechanisms due to collision with the residual gas in the vacuum chamber, black-body radiation, emission, scattering, and absorption, taking into account the heating of the particle (photonic environment) during the trapping period and the subsequent cooling during free-fall. Moreover, we take into account the scattering and absorption of grating photons by employing the formalism developed in Ref.~\cite{belenchia2019talbot}, i.e., without resorting to Rayleigh approximation. The resulting equation for the interference figure is given by 
\begin{align}\label{pattern}
&\frac{P\left(x\right)}{\delta}= 
1+2\sum_{n=0}^{\infty} R_{n} \,B_{n} \left[\frac{n dt_{2}}{t_{T}D}\right] \cos\left(\frac{2 \pi n x}{D}\right)e^{-2\left(\frac{n\pi\sigma_{x}t_{2}}{Dt_1}\right)^2}.
 \end{align}
 with $\delta=\frac{m}{\sqrt{2 \pi} \sigma_{p}(t_{1}+t_{2})}$. Here, the functions $B_{n}[y]$'s are the generalized Talbot coefficients that account for the coherent and incoherent effects of the grating laser (cf. Ref.~\cite{SM} for their explicit expression and Ref.~\cite{belenchia2019talbot} for their derivation), $d=\lambda/2$ is the grating period, $t_T= md^2/h$ is the so called Talbot time and $D=d(t_1+t_2)/t_1$. The initial thermal state, obtained after the cooling time $t_c$, is characterised by its thermal position and momentum standard deviations, $\sigma_x$ and $\sigma_p$ respectively. 
The effect of environmental decoherence, other than scattering and absorption of grating photons, is accounted for by the kernels $R_{n}$, which describe decoherence due to absorption, emission, and scattering of thermal radiation as well as collisional decoherence due to residual gas~\cite{bateman2014near}, i.e., the most relevant decoherence sources acting during the free-fall times. 

In order to properly include the effect of CMs, we rely on Eq.~\eqref{cslme}. Once specified for a spherical homogeneous particle of radius $R$, this master equation allows to obtain an extra decoherence kernel to be included into the $R_n$'s. Such extra kernel is written as
\begin{align}\label{eq:Rcsl}
R^{\text{CSL}}_{n}= e^{ \Gamma_{\!\text{CSL}}\!\left[f\left(\!\frac{ hnqt_{2}}{m D}\!\right)-1\right](t_{1}+t_{2})},
\end{align}
where we have defined
\begin{align}
\Gamma_{\!\text{CSL}}&= 4 \sqrt{\frac{2}{\pi}}\frac{\lambda\, r_{c}^{3}}{\hbar^{3} m_{0}^{2}} \int dq q^2 e^{-r_{c}^{2}q^2/\hbar^2}\tilde{\mu}(q)^{2},  \nonumber\\
f(x)&=4\sqrt{\frac{2 }{\pi}} \frac{\lambda\,r_{c}^{3}}{\hbar^3\,\Gamma_{\!\text{CSL}}}\int dq q^{2} e^{-r_{c}^{2}q^{2}/\hbar^{2}} \tilde{\mu}(q)^{2} \text{Si}\left(\frac{q x}{\hbar}\right),
\end{align}
and
$\tilde{\mu}(\mathbf{q})= ({4\pi \hbar}/{|\mathbf{q}|}) J_{1}\left(\frac{|\mathbf{q}| R}{\hbar}\right)$
is the Fourier transform of the spherical mass distribution, with SI the sine integral.

\begin{figure}[h!]
\centering
\includegraphics[width=1.0\linewidth]{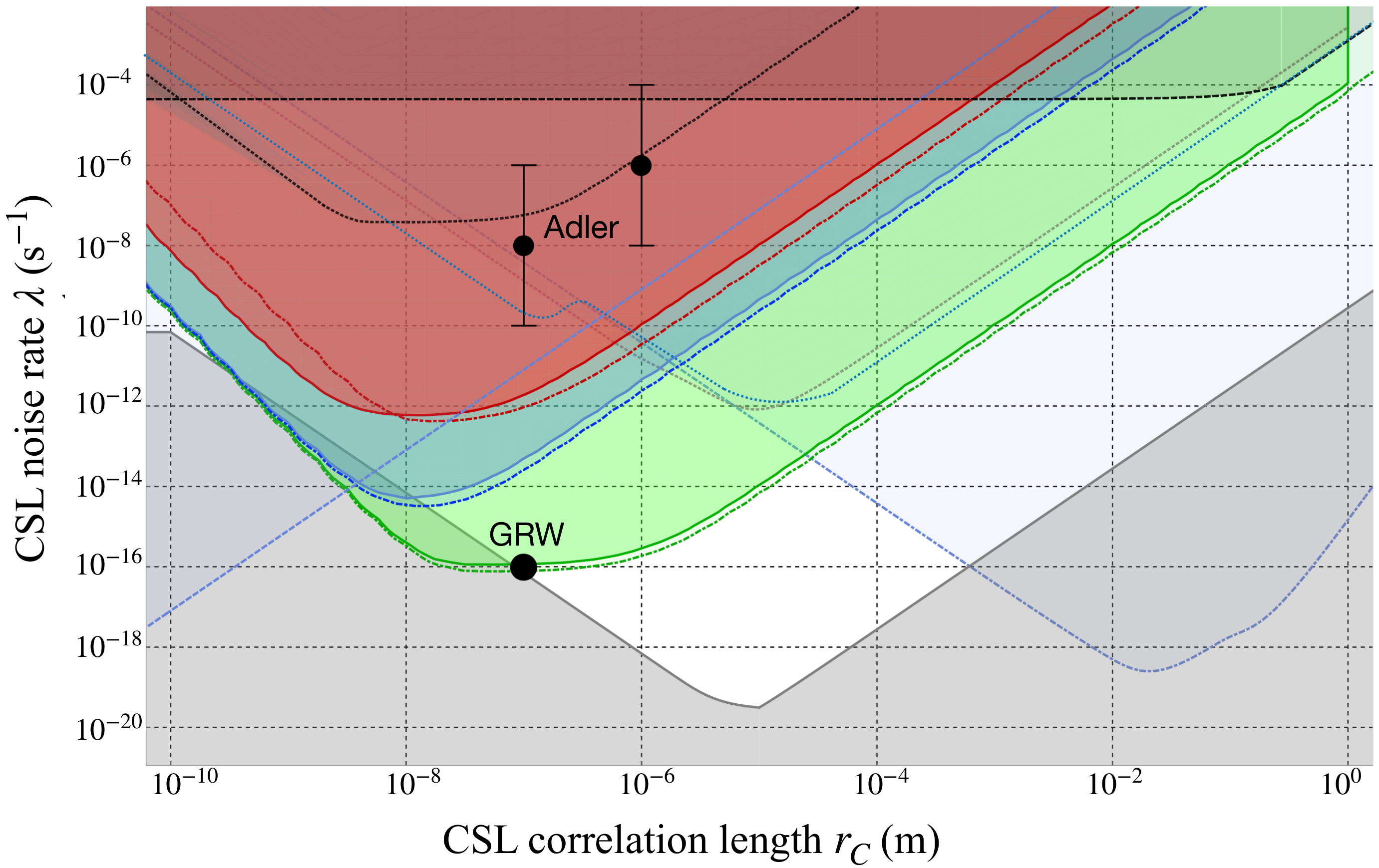}
\caption{ Exclusion plots for the CSL parameters with respect to the GRW’s and Adler’s theoretical values~\cite{ghirardi1986unified,adler2007collapse,adler2007corrigendum}.
The red line denotes the upper bound that can be obtained by the experiment proposed in Ref.~\cite{bateman2014near} with Si particles of mass of $10^6 $~amu at room temperature. The light blue line denotes the upper bound  that can be obtained by using a mass of $10^7$ amu and times $t_{1}=2 t_2$ and  $t_{2}=0.181 s$ at cryogenic temperature. The green line represents the upper bound that can be obtained by using a mass of $10^7$ amu, times $t_{1}=t_2=9.95 s$, and cryogenic temperature.
The dot-dashed lines correspond to the upper bound that can be obtained replacing the Si particle with a SiO2 particle and working at cryogenic temperature. 
 The grey area at the bottom of the plot represents the theoretical bound given in  Refs.~\cite{fein2019quantum,torovs2018bounds,kovachy2015quantum}, the dashed cyan excluded areas represents the bounds given by the present non-interferometric tests~\cite{vinante2020challenging,piscicchia2017csl,carlesso2016experimental,vinante2016upper}, and the dashed black line represents the bounds given by current interferometric tests. }
\label{fig:cslplot}
\end{figure}

In order to analyze the sensitivity of the Talbot--Lau interferometer to CSL noise we define the  following figure of merit
\begin{align}
\aleph= \frac{1}{L}\int_{-L/2}^{L/2}  \frac{|P_{\rm QM}(x)-P_{\rm CSL}(x)|}{|P_{\rm QM}(x)+P_{\rm CSL}(x)|} dx
\end{align}
where $L = 10^{-7}$~m is the spatial window in which the position measurement is performed and $P_{\rm CSL}$ ($P_{\rm QM}$) is the interference figures obtained by assuming the presence (absence) of CSL. Furthermore, we assume in our analysis to be able to discriminate a difference between the two interference pattern bigger than $5\%$, i.e.  
for values of $\aleph \ge 0.05$, which is an experimentally justifiable choice~\cite{juffmann2013experimental}. The results of such analysis are presented in Fig.~\ref{fig:cslplot} from which one can clearly see the great sensitivity of the {single-grating Talbot interferometer for nanoparticle with mass $10^6$~amu as proposed in Ref.~\cite{bateman2014near}} (red shaded area) to CM-based mechanisms. Indeed, from our analysis, this interferometric scheme for $10^6$~amu particles would be able to completely rule out Adler's proposed values for CSL's parameters~\cite{adler2007corrigendum,adler2007lower} and furthermore test CM in an unexplored region of their parameter space. 

{In the next Section, we delve into the experimental feasibility of ground-based near-field interferometric experiments and we show the current limitations -- in terms of size and mass of nanoparticles -- to test collapse models via near-field interferometric setups.}

\section{Experimental perspectives}
\label{perspectives}
The near-field Talbot effect of coherent wave propagation offers the extraordinary advantage of reducing the required level of the initial matter-wave coherence. However, for large particles, the general feasibility of the interferometric scheme, as well as the challenges in terms of technical details are equally important.

{\it Wavefunction evolution, free or not free --} The main challenge for (near-field) interferometric experiments with large particles is to allow for a long enough free evolution time of the prepared quantum superposition state in order to be sensitive to the CMs effects and, even more crucially, to probe the superposition principle of quantum mechanics. The free evolution -- the spatial spreading of the wavefunction $\Psi(r, t)$ with time -- according to the time-dependent Schr\"odinger equation 
\begin{equation}
\frac{\partial}{\partial t}\Psi(r, t) = -i\frac{\hbar}{2m} \nabla^2 \Psi(r, t),
\end{equation}
describes a diffusive process for probability amplitudes similar to a typical diffusion equation with the imaginary diffusion coefficient $-i\hbar/2m$. Therefore, the spreading of $\Psi(r, t)$ scales inverse with particle mass $m$. It should be noted that, for the specific Talbot interferometric scheme considered in this work, there are two free evolution times, $t_1$ and $t_2$, where the first one is necessary for the coherence length of the particle's wave function to cover at least two "slits" of the optical grating. In this setup, considering for instance a $10^8$ amu particle, it would already take so long to show the interference pattern in a matter-wave experiment that the particle would significantly drop in the Earth's gravitational field. 

 
Different solutions are thinkable. One can of course envisage building a 10-100 m fountain, and similar projects are underway for interferometry with ultra-cold atoms. Needless to say, cold atom technologies have matured considerably more than their analogues for manipulation of large-mass particles.
Alternatively, one can consider to levitate the particle by a force field to compensate for the drop in gravity, but here we face the problem of minimizing the additional decoherence effect that such a force field would introduce.
A viable solution is to coherently accelerate the spread of the wavefunction by the action of suitable potentials. The proposals in Refs.~\cite{scala2013matter, romero2017coherent, pino2018chip, bose2017spin} are along these lines.
They use a magnetic field gradient to rapidly separate the two amplitudes of a spin superposition state in space~\cite{bose2017spin}, or alternatively to accelerate the spread of the wave function before and after the splitting, obtained by the action of a double well potential~\cite{pino2018chip,romero2017coherent}.  One of the major obstacles in this endeavour is represented by the additional decoherence introduced by stochastic forces due to vibrations. These ideas are still awaiting their technical realisation for large masses.


Since in the specific setup that we are considering the nanoparticle is optically interacting, we cannot use magnetic gradients as in Ref.~\cite{pino2018chip,bose2017spin} , but we would need to resort to an optical field to increase the spread of the wave function.
 Also here the main limiting factor is the additional decoherence which would need to be kept under control to avoid losing all quantum interference effects.

Alternatively, a more realistic implementation of the near-field experiment, given current technical capabilities, is to allow for long enough free evolution times by freely fall the whole interferometer apparatus in a co-moving reference frame with the particle. This is the idea of a recent proposal for a dedicated satellite mission in space to perform large-mass matter-wave interference experiments with micro- and nano-particles~\cite{kaltenbaek2016macroscopic}. A further option is to consider to throw upwards and catch the very same particle multiple times and use the fountain trajectory to perform interferometry. For all those options, it becomes obvious that it is required to dramatically change the way in which large-mass matter-wave interferometry experiments are performed -- compared to existing technology~\cite{juffmann2013experimental} -- in order to enable such experiments beyond masses of about $10^7$ amu. 
\\
{\it Table-top interferometric experimental tests of collapse models --} Let us now consider again the experiment proposed in~\cite{bateman2014near}. In particular, we consider the same material parameters as in Table~\ref{tab:parameters} but allow for different free falling times $t_1$ and $t_2$ and different masses. 
In order to estimate the potential of such experiments we use the $\aleph$ figure of merit as discussed before and we refer to Fig.~\ref{fig:cslplot}. In Figure~\ref{fig:cslplot}, we report the exclusion plots obtained by simulating the experiment with Si and SiO2 nanoparticles having masses of $10^6$ and $10^7$~amu. It should be noted that, for masses of the order of $10^7$~amu, and for the specifics of the experiment that we are considering, $2\pi R/\lambda\sim 0.2$ which clearly does not allow  to use the point-particle approximation. As already discussed, this is not an issue for our simulations since we adopt the formalism developed in~\cite{belenchia2019talbot}. These results are thus the first comprehensive account of the potential of near-field interferometry with large masses beyond the Rayleigh approximation for fundamental physics studies.

The red-shaded area in Fig.~\ref{fig:cslplot} is the result obtained by using Si particles with mass $10^6$~amu and shows how a table-top, room temperature experiment like the one proposed in~\cite{bateman2014near} could rule out a part of the parameter space including Adler's values of the free parameters for the CSL. It should be noted that, this result is comparable with the very recent one of Ref.~\cite{schrinski2020rule} where the authors considered bounds on CSL free parameters coming from BEC interferometry. In particular, the red-shaded region in Fig.~\ref{fig:cslplot} is comparable with the exclusion plots that Ref.~\cite{schrinski2020rule} argue to be achievable with state-of-the-art BEC interferometry.
The light-blue shaded area shows the improvement that can be obtained increasing the mass to $10^7$~amu, overpassing the results obtained from X-ray emission from a Germanium sample~\cite{piscicchia2017csl} (left dashed line).
In this case, we assume the experiment to be performed in a cryogenic environment at 4~K. This is due to the fact that, at room temperature, increasing the mass (and so the size) of the particles leads to an increase in environmental decoherence affecting the performances of the experiment. In the case we are considering, $t_1 =0.362$~s 
corresponding to the time needed for the wave function to cover exactly four slits~\footnote{
The state of the particle when it is released from the optical trap can be model, to a good approximation, by a thermal harmonic oscillator state (see also Supplemental in~\cite{bateman2014near}).
Assuming $\sigma_{x}$ to be the spatial spread of the state just after the particle is release from the optical trap, after a time $\tau$ the spread, due to the free evolution, will be $
\sigma_{x}(\tau)= \sqrt{\sigma_{x}+\frac{\hbar^{2} \tau^2 }{4 m^{2} \sigma_{x}} }. $
Knowing that to cover n slits of width d, we need $\sigma_{x}(\tau)\ge n d$, and recalling that in our set-up $\sigma_{x}\ll d$ we end up with $
\tau \ge \frac{2 n d}{\hbar m \sigma_{x}}     $
or equivalently $
\tau \ge \frac{4\pi n t_{T}}{d \sigma_{x}},      
$ where $t_{T}$ represents the Talbot time.
} and $t_2=t_1/2$  which implies a drop length of around 1.5~m  
 making the experiment not excessively demanding. 

\begin{figure*}
\centering
\includegraphics[width=1.0\linewidth]{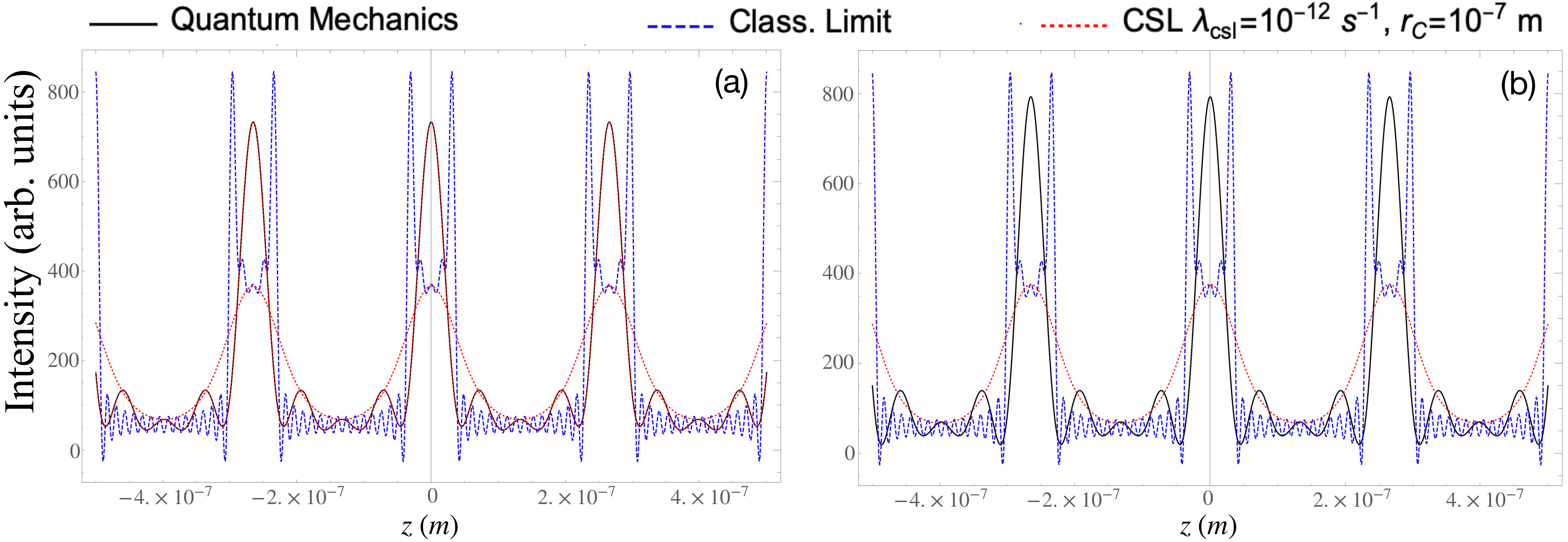}
\caption{Interference pattern for the proposed near field interferometric experiments with nanoparticle of mass $10^7$~amu. Panel \textbf{(a)} shows the interference pattern obtained using Si particles at cryogenic temperature $T_{env}=4K$ and with $E_L/a_L=0.003$~${\rm J\,m^{-2}}$. Panel \textbf{(b)} shows instead the analogue result for SiO2 particles, with $E_L/a_L=0.009$~${\rm J\,m^{-2}}$. In both cases the free fall times are such that $t_1=2t_2=0.36$~s and the pulse energy to spot-area ratio is determined at the same value of the eikonal phase $\phi_0=1.4\pi$ (cf. Refs~\cite{bateman2014near, belenchia2019talbot} for further details on the connection between the eikonal phase and the energy to spot-area ratio). The blue-dashed curves represent the ``classical'' shadow effect image, i.e., the figure that would be obtained from the deviation of the ballistic trajectory of classical particles; the black-continuous curves represent the prediction of standard quantum mechanics; and the red-dotted curves show the effect of accounting for CSL with $\lambda=10^{-12}$~s$^{-1}$ and $r_C=10^{-7}$~m.}
\label{fig:SiSiO2}
\end{figure*}

One can finally wonder what would happen by allowing longer times for the wave-function spreading in the experiment. It is indeed encouraging that in the field of atom interferometry it has been recently shown that optical suspension can be maintained for 20~s coherently~\cite{xu2019probing}. This is a new experimental result which was not thought to be easily achievable beforehand and justify further investigation towards extending such results to large dielectric particles. Motivated by these results, let us then assume an overall wave-function spread time of 20~s for a $10^7$~amu Si particle in a cryogenic environment at 4~K. Based on our analysis, this would suffice to achieve the results shown by the light-green shaded area in Fig.~\ref{fig:SiSiO2}. This shows that a large section of the CSL parameter space could be ruled out, even arriving to rule out the extremely challenging to probe values proposed by Ghirardi Rimini and Weber (GRW)~\cite{ghirardi1986unified}. Once again, this result is comparable with the ones shown in~\cite{schrinski2020rule} for the ultimate test with BEC interferometry.

Our results apply, strictly speaking, to the case in which the nanoparticles are in free fall. However, in the Earth's gravitational field, a free fall time of 20~s would correspond to a drop length of over a kilometer. One possible solution is to move the experiment in space~\cite{kaltenbaek2016macroscopic}, where there is the realistic possibility to achieved free falling times of the order of 100~s and push the mass limit up to $10^{11}$~amu. 

Alternatively, one could consider to somehow suspending the particles, drawing inspiration from  the results in~\cite{xu2019probing}. This could be achieved by optically levitating the particles in a wide trap which is not confining/constraining the free evolution of the wave-function on the order of the size of the trapped particle. As discussed before, the main limitation here would be given by the additional decoherence effects introduced by the scattering and absorption of the trap laser photons.
To circumvent such problems, a possible alternative is the use of electric or magnetic field configurations to compensate the gravitational force which would thus allow to overcome most of the decoherence and heating problems of a standard optical trap~\cite{pino2018chip, scala2013matter}. Nevertheless, a detailed analysis of an experiment with suspended particles would require to properly account for the extra decoherence introduced by such mechanisms, a point that goes beyond the scopes of the current work.



{\it Technical challenges --}
In order to reduce the decohering effects of gas collisions, experiments have to be performed in ultra-high vacuum. In our simulations, we have modelled collisional decoherence by assuming the gas pressure reported in Table~\ref{tab:parameters}. Such vacuum is routine in surface science experiments and considerably higher vacuum conditions have reportedly been achieved by cryogenic ion trapping experiments, initially for antimatter~\cite{gabrielse1990thousandfold}, and currently routinely in atomic ion traps~\cite{schwarz2012cryogenic}. 

In addition to ultra-high vacuum conditions, the use of cryogenic technology has the advantage of automatically reducing all of the thermal radiation decoherence effects as all parts of the experiment and the particle itself will eventually thermalise to low temperature. In our simulations, decoherence effects related to thermal radiation have been modelled for both a room temperature environment and in the case of environmental temperature around 4~K, which is typical in experiments cooled by liquid helium. However, in the considered set-up, the strongest heating effect is due to the absorption of photons from the optical trapping laser.\footnote{
{
Note that a Si/SiO2 particle at cryogenic temperature, can easily reach internal temperature above 200~K if exposed for 1~s to light at wavelength of 1550~nm focused with a $0.8 NA$ lens with intensity $I = 90$~mW/$\mu$m and reach internal temperature of more than 500~K  if at room temperature.}}
The related decoherence processes, due to thermal photon emission, have been estimated based on absorption cross section for the materials of the particle proposed for use. 
{Our study shows that decoherence effects due to thermal photon emission are almost negligible in a $20$~s experiment at cryogenic temperature but relevant in a room temperature environment~\footnote{We have estimated a reduction of the visibility less than $0.0002 \%$ due to the presence of thermal photon emission decoherence for both Si and SiO2 particles with mass of $10^7$~amu in the experimental setup describe by the parameters in Tab.~\ref{tab:parameters} with the cryogenic environmental temperature of $4$~K.}. This suggests that a good strategy to mitigate photon emission decoherence could be to work at cryogenic temperature, if a not too long preparation time is required.}

While silica and silicon particles are the work horse in all levitated optomechanics experiments, and absorption is small for such particles, the particle absorption challenge can be further addressed by material science into reducing absorption at the trapping laser wavelength or by using a refrigerating scheme able to reduce the effects of the absorption during the trapping time and keep the internal particle temperarature low
or, more radically, by removing all lasers from the interferometric scheme and start with a non-optical trap, e.g. a magnetic trap as recently proposed~\cite{pino2018chip}. 


Particle counting statistics represents a further challenge, as experiments based on trapped nanoparticles are single-particle experiments the interferometer sequence has to be repeated many time -- at least 1,000 to 10,000 times to achieve particle number or counting statistics in form of an interference pattern showing the typical structure of delocalised states. Beside the option to consider multiple particle traps to start with , and the promises of alternative quantum figures of merit such as by dynamical model or hypothesis testing~\cite{ralph2018dynamical, schrinski2019macroscopicity}, 
the most pragmatic approach might be to consider a fast reloading or recycling technique to use the same particle (throw and catch)~\cite{hebestreit2018sensing} over and over again. Another option to recycle the same particle would be a co-moving interferometer in a one meter high Einstein elevator type setup. Such an experimental platform has been achieved recently~\cite{bongs2019taking} to test cold atom interferometers in a 0-g simulator in the lab, but it is a serious engineering task to develop a stable elevator which simulates an almost perfect parabolic up-and-down motion. 

{\it Material challenges --} Another benefit of using the same particle for multiple experimental runs is that this would reduce the spread in particle parameters such as mass/size, chemical composition, electrical charges and optical properties (polarizabilties, moments and dielectric function, and the refractive index). As worked out in detail recently, large particles have non-trivial interaction with the grating laser standing-wave~\cite{belenchia2019talbot}. If the particle was used over and over again, then the ensemble would be one of truly mono-disperse particles. Nonetheless, the Talbot near-field scheme is known to tolerate dispersion in de Broglie wavelength of the test particles, and therefore mass, to some degree: $\frac{\Delta\lambda_{dB}}{\lambda_{dB}}=15\%$~\cite{juffmann2013experimental}, assuming a small (longitudinal and transverse) velocity distributions after initial cooling and equal drop distance. Thus, some particle related challenges can be handled by the interferometer itself. Another factor to be controlled is the electrical charge of the particles. Ideally a zero net charge should to be achieved. Multiple studies have investigated the charge challenge and solutions exist~\cite{hempston2017force, frimmer2017controlling}. 
Finally, recent studies show that silica particles are not stable in the optical trap, where they experience a large laser intensity, but see a change of their mass density. This has been attributed to the evaporation of water out of the structure of the nanoparticle originated in the particle synthesis~\cite{hebestreit2017thermal}. While the light-matter interaction of dielectric nanoparticles remains an interesting field of scientific research with many effects yet to be discovered, already acquired knowledge about nanoparticles optical properties gives reason to be optimistic to use them as candidates for matter-wave interferometry experiments such a the ones discussed here.


\section{Conclusion}
\label{conc}
We have shown the possibility for a ground-based Talbot interferometric scheme with large particles to directly test CMs and the validity of the superposition principle at the large-scale limit. In particular, with new technological advancements, this kind of direct test is becoming competitive to the most used, indirect, non-interferometric experiments and comparable to very recent advancements in BEC interferometry~\cite{schrinski2020rule}.


While we have shown that feasible table-top experiments with $10^7$ amu particles in cryogenic environments promise to put stringent constraints on the CSL parameter space, the challenges and possible solutions discussed aim to use technologies fit for an even larger leap in mass. We have shown that there is a window of opportunity for testing a large part of the CSL parameter space down to the challenging historical values of GRW. Indeed, the technology of (levitated) optomechanics has progressed very well in the past ten years or so, that the Talbot schemes can be realistically approached now.

In conclusion, the multiple challenges addressed in this work lead to the realization that increasing the extent of the wavefunction beyond the size of the particle, namely at around 100 nm, is a formidable research challenge which is however possible to tackle with present day technology. We hope that these results will motivate the theoretical and experimental community to engage in a joint effort to further investigate these  architectures in the near future.


\acknowledgments
The authors acknowledge support from the H2020-FETOPEN-2018-2020  project  TEQ  (grantnr. 766900). AB acknowledges support from the MSCA  IF  project  pERFEcTO  (grant  nr.795782). GG and HU acknowledge support from the Leverhulme Trust (RPG-2016-046).
GG also acknowledge support from the Spanish  Agencia Estatal de Investigación, project PID2019-107609GB-I00/AEI / 10.13039/501100011033.
MP acknowledges support from the DfE-SFI Investigator Programme (grant 15/IA/2864), COST Action CA15220, the Royal Society Wolfson Research Fellowship (RSWF\textbackslash R3\textbackslash183013), the Royal Society International Exchanges Programme (IEC\textbackslash R2\textbackslash192220), the Leverhulme Trust Research Project Grant (grant nr.~RGP-2018-266), and the UK EPSRC.


\bibliography{references}

\begin{thebibliography}{2}
\expandafter\ifx\csname natexlab\endcsname\relax\def\natexlab#1{#1}\fi
\expandafter\ifx\csname bibnamefont\endcsname\relax
  \def\bibnamefont#1{#1}\fi
\expandafter\ifx\csname bibfnamefont\endcsname\relax
  \def\bibfnamefont#1{#1}\fi
\expandafter\ifx\csname citenamefont\endcsname\relax
  \def\citenamefont#1{#1}\fi
\expandafter\ifx\csname url\endcsname\relax
  \def\url#1{\texttt{#1}}\fi
\expandafter\ifx\csname urlprefix\endcsname\relax\def\urlprefix{URL }\fi
\providecommand{\bibinfo}[2]{#2}
\providecommand{\eprint}[2][]{\url{#2}}

\bibitem[{\citenamefont{Bateman et~al.}(2014)\citenamefont{Bateman,
  Nimmrichter, Hornberger, and Ulbricht}}]{bateman2014near}
\bibinfo{author}{\bibfnamefont{J.}~\bibnamefont{Bateman}},
  \bibinfo{author}{\bibfnamefont{S.}~\bibnamefont{Nimmrichter}},
  \bibinfo{author}{\bibfnamefont{K.}~\bibnamefont{Hornberger}},
  \bibnamefont{and} \bibinfo{author}{\bibfnamefont{H.}~\bibnamefont{Ulbricht}},
  \bibinfo{journal}{Nat. Commun.} \textbf{\bibinfo{volume}{5}},
  \bibinfo{pages}{1} (\bibinfo{year}{2014}).

\bibitem[{\citenamefont{Belenchia et~al.}(2019)\citenamefont{Belenchia,
  Gasbarri, Kaltenbaek, Ulbricht, and Paternostro}}]{belenchia2019talbot}
\bibinfo{author}{\bibfnamefont{A.}~\bibnamefont{Belenchia}},
  \bibinfo{author}{\bibfnamefont{G.}~\bibnamefont{Gasbarri}},
  \bibinfo{author}{\bibfnamefont{R.}~\bibnamefont{Kaltenbaek}},
  \bibinfo{author}{\bibfnamefont{H.}~\bibnamefont{Ulbricht}}, \bibnamefont{and}
  \bibinfo{author}{\bibfnamefont{M.}~\bibnamefont{Paternostro}},
  \bibinfo{journal}{Physical Review A} \textbf{\bibinfo{volume}{100}},
  \bibinfo{pages}{033813} (\bibinfo{year}{2019}).

\end{thebibliography}


\begin{thebibliography}{49}%
\makeatletter
\providecommand \@ifxundefined [1]{%
 \@ifx{#1\undefined}
}%
\providecommand \@ifnum [1]{%
 \ifnum #1\expandafter \@firstoftwo
 \else \expandafter \@secondoftwo
 \fi
}%
\providecommand \@ifx [1]{%
 \ifx #1\expandafter \@firstoftwo
 \else \expandafter \@secondoftwo
 \fi
}%
\providecommand \natexlab [1]{#1}%
\providecommand \enquote  [1]{``#1''}%
\providecommand \bibnamefont  [1]{#1}%
\providecommand \bibfnamefont [1]{#1}%
\providecommand \citenamefont [1]{#1}%
\providecommand \href@noop [0]{\@secondoftwo}%
\providecommand \href [0]{\begingroup \@sanitize@url \@href}%
\providecommand \@href[1]{\@@startlink{#1}\@@href}%
\providecommand \@@href[1]{\endgroup#1\@@endlink}%
\providecommand \@sanitize@url [0]{\catcode `\\12\catcode `\$12\catcode
  `\&12\catcode `\#12\catcode `\^12\catcode `\_12\catcode `\%12\relax}%
\providecommand \@@startlink[1]{}%
\providecommand \@@endlink[0]{}%
\providecommand \url  [0]{\begingroup\@sanitize@url \@url }%
\providecommand \@url [1]{\endgroup\@href {#1}{\urlprefix }}%
\providecommand \urlprefix  [0]{URL }%
\providecommand \Eprint [0]{\href }%
\providecommand \doibase [0]{http://dx.doi.org/}%
\providecommand \selectlanguage [0]{\@gobble}%
\providecommand \bibinfo  [0]{\@secondoftwo}%
\providecommand \bibfield  [0]{\@secondoftwo}%
\providecommand \translation [1]{[#1]}%
\providecommand \BibitemOpen [0]{}%
\providecommand \bibitemStop [0]{}%
\providecommand \bibitemNoStop [0]{.\EOS\space}%
\providecommand \EOS [0]{\spacefactor3000\relax}%
\providecommand \BibitemShut  [1]{\csname bibitem#1\endcsname}%
\let\auto@bib@innerbib\@empty
\bibitem [{\citenamefont {Joos}\ and\ \citenamefont
  {Zeh}(1985)}]{joos1985emergence}%
  \BibitemOpen
  \bibfield  {author} {\bibinfo {author} {\bibfnamefont {E.}~\bibnamefont
  {Joos}}\ and\ \bibinfo {author} {\bibfnamefont {H.~D.}\ \bibnamefont {Zeh}},\
  }\href@noop {} {\bibfield  {journal} {\bibinfo  {journal} {Zeitschrift
  f{\"u}r Physik B Condensed Matter}\ }\textbf {\bibinfo {volume} {59}},\
  \bibinfo {pages} {223} (\bibinfo {year} {1985})}\BibitemShut {NoStop}%
\bibitem [{\citenamefont {Caldeira}\ and\ \citenamefont
  {Leggett}(1983)}]{caldeira1983path}%
  \BibitemOpen
  \bibfield  {author} {\bibinfo {author} {\bibfnamefont {A.~O.}\ \bibnamefont
  {Caldeira}}\ and\ \bibinfo {author} {\bibfnamefont {A.~J.}\ \bibnamefont
  {Leggett}},\ }\href@noop {} {\bibfield  {journal} {\bibinfo  {journal}
  {Physica A: Statistical mechanics and its Applications}\ }\textbf {\bibinfo
  {volume} {121}},\ \bibinfo {pages} {587} (\bibinfo {year}
  {1983})}\BibitemShut {NoStop}%
\bibitem [{\citenamefont {Zurek}(1982)}]{zurek1982environment}%
  \BibitemOpen
  \bibfield  {author} {\bibinfo {author} {\bibfnamefont {W.~H.}\ \bibnamefont
  {Zurek}},\ }\href@noop {} {\bibfield  {journal} {\bibinfo  {journal}
  {Physical review D}\ }\textbf {\bibinfo {volume} {26}},\ \bibinfo {pages}
  {1862} (\bibinfo {year} {1982})}\BibitemShut {NoStop}%
\bibitem [{\citenamefont {Zurek}(1981)}]{zurek1981pointer}%
  \BibitemOpen
  \bibfield  {author} {\bibinfo {author} {\bibfnamefont {W.~H.}\ \bibnamefont
  {Zurek}},\ }\href@noop {} {\bibfield  {journal} {\bibinfo  {journal}
  {Physical review D}\ }\textbf {\bibinfo {volume} {24}},\ \bibinfo {pages}
  {1516} (\bibinfo {year} {1981})}\BibitemShut {NoStop}%
\bibitem [{\citenamefont {Zeh}(1973)}]{zeh1973toward}%
  \BibitemOpen
  \bibfield  {author} {\bibinfo {author} {\bibfnamefont {H.~D.}\ \bibnamefont
  {Zeh}},\ }\href@noop {} {\bibfield  {journal} {\bibinfo  {journal}
  {Foundations of Physics}\ }\textbf {\bibinfo {volume} {3}},\ \bibinfo {pages}
  {109} (\bibinfo {year} {1973})}\BibitemShut {NoStop}%
\bibitem [{\citenamefont {Zeh}(1970)}]{zeh1970interpretation}%
  \BibitemOpen
  \bibfield  {author} {\bibinfo {author} {\bibfnamefont {H.~D.}\ \bibnamefont
  {Zeh}},\ }\href@noop {} {\bibfield  {journal} {\bibinfo  {journal}
  {Foundations of Physics}\ }\textbf {\bibinfo {volume} {1}},\ \bibinfo {pages}
  {69} (\bibinfo {year} {1970})}\BibitemShut {NoStop}%
\bibitem [{\citenamefont {Bassi}\ and\ \citenamefont
  {Ghirardi}(2003)}]{bassi2003dynamical}%
  \BibitemOpen
  \bibfield  {author} {\bibinfo {author} {\bibfnamefont {A.}~\bibnamefont
  {Bassi}}\ and\ \bibinfo {author} {\bibfnamefont {G.}~\bibnamefont
  {Ghirardi}},\ }\href@noop {} {\bibfield  {journal} {\bibinfo  {journal}
  {Physics Reports}\ }\textbf {\bibinfo {volume} {379}},\ \bibinfo {pages}
  {257} (\bibinfo {year} {2003})}\BibitemShut {NoStop}%
\bibitem [{\citenamefont {Ghirardi}\ \emph {et~al.}(1990)\citenamefont
  {Ghirardi}, \citenamefont {Pearle},\ and\ \citenamefont
  {Rimini}}]{ghirardi1990markov}%
  \BibitemOpen
  \bibfield  {author} {\bibinfo {author} {\bibfnamefont {G.~C.}\ \bibnamefont
  {Ghirardi}}, \bibinfo {author} {\bibfnamefont {P.}~\bibnamefont {Pearle}}, \
  and\ \bibinfo {author} {\bibfnamefont {A.}~\bibnamefont {Rimini}},\
  }\href@noop {} {\bibfield  {journal} {\bibinfo  {journal} {Physical Review
  A}\ }\textbf {\bibinfo {volume} {42}},\ \bibinfo {pages} {78} (\bibinfo
  {year} {1990})}\BibitemShut {NoStop}%
\bibitem [{\citenamefont {Ghirardi}\ \emph {et~al.}(1986)\citenamefont
  {Ghirardi}, \citenamefont {Rimini},\ and\ \citenamefont
  {Weber}}]{ghirardi1986unified}%
  \BibitemOpen
  \bibfield  {author} {\bibinfo {author} {\bibfnamefont {G.~C.}\ \bibnamefont
  {Ghirardi}}, \bibinfo {author} {\bibfnamefont {A.}~\bibnamefont {Rimini}}, \
  and\ \bibinfo {author} {\bibfnamefont {T.}~\bibnamefont {Weber}},\
  }\href@noop {} {\bibfield  {journal} {\bibinfo  {journal} {Physical review
  D}\ }\textbf {\bibinfo {volume} {34}},\ \bibinfo {pages} {470} (\bibinfo
  {year} {1986})}\BibitemShut {NoStop}%
\bibitem [{\citenamefont {Carlesso}\ and\ \citenamefont
  {Donadi}(2019)}]{carlesso2019collapse}%
  \BibitemOpen
  \bibfield  {author} {\bibinfo {author} {\bibfnamefont {M.}~\bibnamefont
  {Carlesso}}\ and\ \bibinfo {author} {\bibfnamefont {S.}~\bibnamefont
  {Donadi}},\ }in\ \href@noop {} {\emph {\bibinfo {booktitle} {Advances in Open
  Systems and Fundamental Tests of Quantum Mechanics}}}\ (\bibinfo  {publisher}
  {Springer},\ \bibinfo {year} {2019})\ pp.\ \bibinfo {pages}
  {1--13}\BibitemShut {NoStop}%
\bibitem [{\citenamefont {Fein}\ \emph {et~al.}(2019)\citenamefont {Fein},
  \citenamefont {Geyer}, \citenamefont {Zwick}, \citenamefont {Kia{\l}ka},
  \citenamefont {Pedalino}, \citenamefont {Mayor}, \citenamefont {Gerlich},\
  and\ \citenamefont {Arndt}}]{fein2019quantum}%
  \BibitemOpen
  \bibfield  {author} {\bibinfo {author} {\bibfnamefont {Y.~Y.}\ \bibnamefont
  {Fein}}, \bibinfo {author} {\bibfnamefont {P.}~\bibnamefont {Geyer}},
  \bibinfo {author} {\bibfnamefont {P.}~\bibnamefont {Zwick}}, \bibinfo
  {author} {\bibfnamefont {F.}~\bibnamefont {Kia{\l}ka}}, \bibinfo {author}
  {\bibfnamefont {S.}~\bibnamefont {Pedalino}}, \bibinfo {author}
  {\bibfnamefont {M.}~\bibnamefont {Mayor}}, \bibinfo {author} {\bibfnamefont
  {S.}~\bibnamefont {Gerlich}}, \ and\ \bibinfo {author} {\bibfnamefont
  {M.}~\bibnamefont {Arndt}},\ }\href@noop {} {\bibfield  {journal} {\bibinfo
  {journal} {Nature Physics}\ }\textbf {\bibinfo {volume} {15}},\ \bibinfo
  {pages} {1242} (\bibinfo {year} {2019})}\BibitemShut {NoStop}%
\bibitem [{\citenamefont {Gerlich}\ \emph {et~al.}(2007)\citenamefont
  {Gerlich}, \citenamefont {Hackerm{\"u}ller}, \citenamefont {Hornberger},
  \citenamefont {Stibor}, \citenamefont {Ulbricht}, \citenamefont {Gring},
  \citenamefont {Goldfarb}, \citenamefont {Savas}, \citenamefont {M{\"u}ri},
  \citenamefont {Mayor} \emph {et~al.}}]{gerlich2007kapitza}%
  \BibitemOpen
  \bibfield  {author} {\bibinfo {author} {\bibfnamefont {S.}~\bibnamefont
  {Gerlich}}, \bibinfo {author} {\bibfnamefont {L.}~\bibnamefont
  {Hackerm{\"u}ller}}, \bibinfo {author} {\bibfnamefont {K.}~\bibnamefont
  {Hornberger}}, \bibinfo {author} {\bibfnamefont {A.}~\bibnamefont {Stibor}},
  \bibinfo {author} {\bibfnamefont {H.}~\bibnamefont {Ulbricht}}, \bibinfo
  {author} {\bibfnamefont {M.}~\bibnamefont {Gring}}, \bibinfo {author}
  {\bibfnamefont {F.}~\bibnamefont {Goldfarb}}, \bibinfo {author}
  {\bibfnamefont {T.}~\bibnamefont {Savas}}, \bibinfo {author} {\bibfnamefont
  {M.}~\bibnamefont {M{\"u}ri}}, \bibinfo {author} {\bibfnamefont
  {M.}~\bibnamefont {Mayor}},  \emph {et~al.},\ }\href@noop {} {\bibfield
  {journal} {\bibinfo  {journal} {Nature Physics}\ }\textbf {\bibinfo {volume}
  {3}},\ \bibinfo {pages} {711} (\bibinfo {year} {2007})}\BibitemShut {NoStop}%
\bibitem [{\citenamefont {Brezger}\ \emph {et~al.}(2003)\citenamefont
  {Brezger}, \citenamefont {Arndt},\ and\ \citenamefont
  {Zeilinger}}]{brezger2003concepts}%
  \BibitemOpen
  \bibfield  {author} {\bibinfo {author} {\bibfnamefont {B.}~\bibnamefont
  {Brezger}}, \bibinfo {author} {\bibfnamefont {M.}~\bibnamefont {Arndt}}, \
  and\ \bibinfo {author} {\bibfnamefont {A.}~\bibnamefont {Zeilinger}},\
  }\href@noop {} {\bibfield  {journal} {\bibinfo  {journal} {Journal of Optics
  B: Quantum and Semiclassical Optics}\ }\textbf {\bibinfo {volume} {5}},\
  \bibinfo {pages} {S82} (\bibinfo {year} {2003})}\BibitemShut {NoStop}%
\bibitem [{\citenamefont {Bateman}\ \emph {et~al.}(2014)\citenamefont
  {Bateman}, \citenamefont {Nimmrichter}, \citenamefont {Hornberger},\ and\
  \citenamefont {Ulbricht}}]{bateman2014near}%
  \BibitemOpen
  \bibfield  {author} {\bibinfo {author} {\bibfnamefont {J.}~\bibnamefont
  {Bateman}}, \bibinfo {author} {\bibfnamefont {S.}~\bibnamefont
  {Nimmrichter}}, \bibinfo {author} {\bibfnamefont {K.}~\bibnamefont
  {Hornberger}}, \ and\ \bibinfo {author} {\bibfnamefont {H.}~\bibnamefont
  {Ulbricht}},\ }\href@noop {} {\bibfield  {journal} {\bibinfo  {journal} {Nat.
  Commun.}\ }\textbf {\bibinfo {volume} {5}},\ \bibinfo {pages} {1} (\bibinfo
  {year} {2014})}\BibitemShut {NoStop}%
\bibitem [{\citenamefont {Kaltenbaek}\ \emph {et~al.}(2016)\citenamefont
  {Kaltenbaek}, \citenamefont {Aspelmeyer}, \citenamefont {Barker},
  \citenamefont {Bassi}, \citenamefont {Bateman}, \citenamefont {Bongs},
  \citenamefont {Bose}, \citenamefont {Braxmaier}, \citenamefont {Brukner},
  \citenamefont {Christophe} \emph {et~al.}}]{kaltenbaek2016macroscopic}%
  \BibitemOpen
  \bibfield  {author} {\bibinfo {author} {\bibfnamefont {R.}~\bibnamefont
  {Kaltenbaek}}, \bibinfo {author} {\bibfnamefont {M.}~\bibnamefont
  {Aspelmeyer}}, \bibinfo {author} {\bibfnamefont {P.~F.}\ \bibnamefont
  {Barker}}, \bibinfo {author} {\bibfnamefont {A.}~\bibnamefont {Bassi}},
  \bibinfo {author} {\bibfnamefont {J.}~\bibnamefont {Bateman}}, \bibinfo
  {author} {\bibfnamefont {K.}~\bibnamefont {Bongs}}, \bibinfo {author}
  {\bibfnamefont {S.}~\bibnamefont {Bose}}, \bibinfo {author} {\bibfnamefont
  {C.}~\bibnamefont {Braxmaier}}, \bibinfo {author} {\bibfnamefont
  {{\v{C}}.}~\bibnamefont {Brukner}}, \bibinfo {author} {\bibfnamefont
  {B.}~\bibnamefont {Christophe}},  \emph {et~al.},\ }\href@noop {} {\bibfield
  {journal} {\bibinfo  {journal} {EPJ Quantum Technology}\ }\textbf {\bibinfo
  {volume} {3}},\ \bibinfo {pages} {5} (\bibinfo {year} {2016})}\BibitemShut
  {NoStop}%
\bibitem [{\citenamefont {Belenchia}\ \emph {et~al.}(2019)\citenamefont
  {Belenchia}, \citenamefont {Gasbarri}, \citenamefont {Kaltenbaek},
  \citenamefont {Ulbricht},\ and\ \citenamefont
  {Paternostro}}]{belenchia2019talbot}%
  \BibitemOpen
  \bibfield  {author} {\bibinfo {author} {\bibfnamefont {A.}~\bibnamefont
  {Belenchia}}, \bibinfo {author} {\bibfnamefont {G.}~\bibnamefont {Gasbarri}},
  \bibinfo {author} {\bibfnamefont {R.}~\bibnamefont {Kaltenbaek}}, \bibinfo
  {author} {\bibfnamefont {H.}~\bibnamefont {Ulbricht}}, \ and\ \bibinfo
  {author} {\bibfnamefont {M.}~\bibnamefont {Paternostro}},\ }\href@noop {}
  {\bibfield  {journal} {\bibinfo  {journal} {Physical Review A}\ }\textbf
  {\bibinfo {volume} {100}},\ \bibinfo {pages} {033813} (\bibinfo {year}
  {2019})}\BibitemShut {NoStop}%
\bibitem [{\citenamefont {Nimmrichter}(2014)}]{nimmrichter2014macroscopic}%
  \BibitemOpen
  \bibfield  {author} {\bibinfo {author} {\bibfnamefont {S.}~\bibnamefont
  {Nimmrichter}},\ }\href@noop {} {\emph {\bibinfo {title} {Macroscopic matter
  wave interferometry}}}\ (\bibinfo  {publisher} {Springer},\ \bibinfo {year}
  {2014})\BibitemShut {NoStop}%
\bibitem [{\citenamefont {Gasbarri}\ \emph {et~al.}(2017)\citenamefont
  {Gasbarri}, \citenamefont {Toro{\v{s}}}, \citenamefont {Donadi},\ and\
  \citenamefont {Bassi}}]{gasbarri2017gravity}%
  \BibitemOpen
  \bibfield  {author} {\bibinfo {author} {\bibfnamefont {G.}~\bibnamefont
  {Gasbarri}}, \bibinfo {author} {\bibfnamefont {M.}~\bibnamefont
  {Toro{\v{s}}}}, \bibinfo {author} {\bibfnamefont {S.}~\bibnamefont {Donadi}},
  \ and\ \bibinfo {author} {\bibfnamefont {A.}~\bibnamefont {Bassi}},\
  }\href@noop {} {\bibfield  {journal} {\bibinfo  {journal} {Physical Review
  D}\ }\textbf {\bibinfo {volume} {96}},\ \bibinfo {pages} {104013} (\bibinfo
  {year} {2017})}\BibitemShut {NoStop}%
\bibitem [{\citenamefont {Ferialdi}\ and\ \citenamefont
  {Bassi}(2012)}]{ferialdi2012dissipative}%
  \BibitemOpen
  \bibfield  {author} {\bibinfo {author} {\bibfnamefont {L.}~\bibnamefont
  {Ferialdi}}\ and\ \bibinfo {author} {\bibfnamefont {A.}~\bibnamefont
  {Bassi}},\ }\href@noop {} {\bibfield  {journal} {\bibinfo  {journal}
  {Physical Review A}\ }\textbf {\bibinfo {volume} {86}},\ \bibinfo {pages}
  {022108} (\bibinfo {year} {2012})}\BibitemShut {NoStop}%
\bibitem [{\citenamefont {Adler}\ and\ \citenamefont
  {Bassi}(2008)}]{adler2008collapse}%
  \BibitemOpen
  \bibfield  {author} {\bibinfo {author} {\bibfnamefont {S.~L.}\ \bibnamefont
  {Adler}}\ and\ \bibinfo {author} {\bibfnamefont {A.}~\bibnamefont {Bassi}},\
  }\href@noop {} {\bibfield  {journal} {\bibinfo  {journal} {Journal of Physics
  A: Mathematical and Theoretical}\ }\textbf {\bibinfo {volume} {41}},\
  \bibinfo {pages} {395308} (\bibinfo {year} {2008})}\BibitemShut {NoStop}%
\bibitem [{\citenamefont {Adler}\ and\ \citenamefont
  {Bassi}(2007)}]{adler2007collapse}%
  \BibitemOpen
  \bibfield  {author} {\bibinfo {author} {\bibfnamefont {S.~L.}\ \bibnamefont
  {Adler}}\ and\ \bibinfo {author} {\bibfnamefont {A.}~\bibnamefont {Bassi}},\
  }\href@noop {} {\bibfield  {journal} {\bibinfo  {journal} {Journal of Physics
  A: Mathematical and Theoretical}\ }\textbf {\bibinfo {volume} {40}},\
  \bibinfo {pages} {15083} (\bibinfo {year} {2007})}\BibitemShut {NoStop}%
\bibitem [{SM()}]{SM}%
  \BibitemOpen
  \href@noop {} {}\bibinfo {note} {Additional details on the analysis reported
  in the main text are reported in the Supplementary Material.}\BibitemShut
  {Stop}%
\bibitem [{\citenamefont {Juffmann}\ \emph {et~al.}(2013)\citenamefont
  {Juffmann}, \citenamefont {Ulbricht},\ and\ \citenamefont
  {Arndt}}]{juffmann2013experimental}%
  \BibitemOpen
  \bibfield  {author} {\bibinfo {author} {\bibfnamefont {T.}~\bibnamefont
  {Juffmann}}, \bibinfo {author} {\bibfnamefont {H.}~\bibnamefont {Ulbricht}},
  \ and\ \bibinfo {author} {\bibfnamefont {M.}~\bibnamefont {Arndt}},\
  }\href@noop {} {\bibfield  {journal} {\bibinfo  {journal} {Reports on
  Progress in Physics}\ }\textbf {\bibinfo {volume} {76}},\ \bibinfo {pages}
  {086402} (\bibinfo {year} {2013})}\BibitemShut {NoStop}%
\bibitem [{\citenamefont {Adler}(2007{\natexlab{a}})}]{adler2007corrigendum}%
  \BibitemOpen
  \bibfield  {author} {\bibinfo {author} {\bibfnamefont {S.~L.}\ \bibnamefont
  {Adler}},\ }\href@noop {} {\bibfield  {journal} {\bibinfo  {journal} {Journal
  of Physics A: Mathematical and Theoretical}\ }\textbf {\bibinfo {volume}
  {40}},\ \bibinfo {pages} {13501} (\bibinfo {year}
  {2007}{\natexlab{a}})}\BibitemShut {NoStop}%
\bibitem [{\citenamefont {Adler}(2007{\natexlab{b}})}]{adler2007lower}%
  \BibitemOpen
  \bibfield  {author} {\bibinfo {author} {\bibfnamefont {S.~L.}\ \bibnamefont
  {Adler}},\ }\href@noop {} {\bibfield  {journal} {\bibinfo  {journal} {Journal
  of Physics A: Mathematical and Theoretical}\ }\textbf {\bibinfo {volume}
  {40}},\ \bibinfo {pages} {2935} (\bibinfo {year}
  {2007}{\natexlab{b}})}\BibitemShut {NoStop}%
\bibitem [{\citenamefont {Toro{\v{s}}}\ and\ \citenamefont
  {Bassi}(2018)}]{torovs2018bounds}%
  \BibitemOpen
  \bibfield  {author} {\bibinfo {author} {\bibfnamefont {M.}~\bibnamefont
  {Toro{\v{s}}}}\ and\ \bibinfo {author} {\bibfnamefont {A.}~\bibnamefont
  {Bassi}},\ }\href@noop {} {\bibfield  {journal} {\bibinfo  {journal} {Journal
  of Physics A: Mathematical and Theoretical}\ }\textbf {\bibinfo {volume}
  {51}},\ \bibinfo {pages} {115302} (\bibinfo {year} {2018})}\BibitemShut
  {NoStop}%
\bibitem [{\citenamefont {Kovachy}\ \emph {et~al.}(2015)\citenamefont
  {Kovachy}, \citenamefont {Asenbaum}, \citenamefont {Overstreet},
  \citenamefont {Donnelly}, \citenamefont {Dickerson}, \citenamefont
  {Sugarbaker}, \citenamefont {Hogan},\ and\ \citenamefont
  {Kasevich}}]{kovachy2015quantum}%
  \BibitemOpen
  \bibfield  {author} {\bibinfo {author} {\bibfnamefont {T.}~\bibnamefont
  {Kovachy}}, \bibinfo {author} {\bibfnamefont {P.}~\bibnamefont {Asenbaum}},
  \bibinfo {author} {\bibfnamefont {C.}~\bibnamefont {Overstreet}}, \bibinfo
  {author} {\bibfnamefont {C.}~\bibnamefont {Donnelly}}, \bibinfo {author}
  {\bibfnamefont {S.}~\bibnamefont {Dickerson}}, \bibinfo {author}
  {\bibfnamefont {A.}~\bibnamefont {Sugarbaker}}, \bibinfo {author}
  {\bibfnamefont {J.}~\bibnamefont {Hogan}}, \ and\ \bibinfo {author}
  {\bibfnamefont {M.}~\bibnamefont {Kasevich}},\ }\href@noop {} {\bibfield
  {journal} {\bibinfo  {journal} {Nature}\ }\textbf {\bibinfo {volume} {528}},\
  \bibinfo {pages} {530} (\bibinfo {year} {2015})}\BibitemShut {NoStop}%
\bibitem [{\citenamefont {Vinante}\ \emph {et~al.}(2020)\citenamefont
  {Vinante}, \citenamefont {Carlesso}, \citenamefont {Bassi}, \citenamefont
  {Chiasera}, \citenamefont {Varas}, \citenamefont {Falferi}, \citenamefont
  {Margesin}, \citenamefont {Mezzena},\ and\ \citenamefont
  {Ulbricht}}]{vinante2020challenging}%
  \BibitemOpen
  \bibfield  {author} {\bibinfo {author} {\bibfnamefont {A.}~\bibnamefont
  {Vinante}}, \bibinfo {author} {\bibfnamefont {M.}~\bibnamefont {Carlesso}},
  \bibinfo {author} {\bibfnamefont {A.}~\bibnamefont {Bassi}}, \bibinfo
  {author} {\bibfnamefont {A.}~\bibnamefont {Chiasera}}, \bibinfo {author}
  {\bibfnamefont {S.}~\bibnamefont {Varas}}, \bibinfo {author} {\bibfnamefont
  {P.}~\bibnamefont {Falferi}}, \bibinfo {author} {\bibfnamefont
  {B.}~\bibnamefont {Margesin}}, \bibinfo {author} {\bibfnamefont
  {R.}~\bibnamefont {Mezzena}}, \ and\ \bibinfo {author} {\bibfnamefont
  {H.}~\bibnamefont {Ulbricht}},\ }\href {\doibase
  10.1103/PhysRevLett.125.100404} {\bibfield  {journal} {\bibinfo  {journal}
  {Phys. Rev. Lett.}\ }\textbf {\bibinfo {volume} {125}},\ \bibinfo {pages}
  {100404} (\bibinfo {year} {2020})}\BibitemShut {NoStop}%
\bibitem [{\citenamefont {Piscicchia}\ \emph {et~al.}(2017)\citenamefont
  {Piscicchia}, \citenamefont {Bassi}, \citenamefont {Curceanu}, \citenamefont
  {Grande}, \citenamefont {Donadi}, \citenamefont {Hiesmayr},\ and\
  \citenamefont {Pichler}}]{piscicchia2017csl}%
  \BibitemOpen
  \bibfield  {author} {\bibinfo {author} {\bibfnamefont {K.}~\bibnamefont
  {Piscicchia}}, \bibinfo {author} {\bibfnamefont {A.}~\bibnamefont {Bassi}},
  \bibinfo {author} {\bibfnamefont {C.}~\bibnamefont {Curceanu}}, \bibinfo
  {author} {\bibfnamefont {R.~D.}\ \bibnamefont {Grande}}, \bibinfo {author}
  {\bibfnamefont {S.}~\bibnamefont {Donadi}}, \bibinfo {author} {\bibfnamefont
  {B.~C.}\ \bibnamefont {Hiesmayr}}, \ and\ \bibinfo {author} {\bibfnamefont
  {A.}~\bibnamefont {Pichler}},\ }\href@noop {} {\bibfield  {journal} {\bibinfo
   {journal} {Entropy}\ }\textbf {\bibinfo {volume} {19}},\ \bibinfo {pages}
  {319} (\bibinfo {year} {2017})}\BibitemShut {NoStop}%
\bibitem [{\citenamefont {Carlesso}\ \emph {et~al.}(2016)\citenamefont
  {Carlesso}, \citenamefont {Bassi}, \citenamefont {Falferi},\ and\
  \citenamefont {Vinante}}]{carlesso2016experimental}%
  \BibitemOpen
  \bibfield  {author} {\bibinfo {author} {\bibfnamefont {M.}~\bibnamefont
  {Carlesso}}, \bibinfo {author} {\bibfnamefont {A.}~\bibnamefont {Bassi}},
  \bibinfo {author} {\bibfnamefont {P.}~\bibnamefont {Falferi}}, \ and\
  \bibinfo {author} {\bibfnamefont {A.}~\bibnamefont {Vinante}},\ }\href@noop
  {} {\bibfield  {journal} {\bibinfo  {journal} {Physical Review D}\ }\textbf
  {\bibinfo {volume} {94}},\ \bibinfo {pages} {124036} (\bibinfo {year}
  {2016})}\BibitemShut {NoStop}%
\bibitem [{\citenamefont {Vinante}\ \emph {et~al.}(2016)\citenamefont
  {Vinante}, \citenamefont {Bahrami}, \citenamefont {Bassi}, \citenamefont
  {Usenko}, \citenamefont {Wijts},\ and\ \citenamefont
  {Oosterkamp}}]{vinante2016upper}%
  \BibitemOpen
  \bibfield  {author} {\bibinfo {author} {\bibfnamefont {A.}~\bibnamefont
  {Vinante}}, \bibinfo {author} {\bibfnamefont {M.}~\bibnamefont {Bahrami}},
  \bibinfo {author} {\bibfnamefont {A.}~\bibnamefont {Bassi}}, \bibinfo
  {author} {\bibfnamefont {O.}~\bibnamefont {Usenko}}, \bibinfo {author}
  {\bibfnamefont {G.}~\bibnamefont {Wijts}}, \ and\ \bibinfo {author}
  {\bibfnamefont {T.}~\bibnamefont {Oosterkamp}},\ }\href@noop {} {\bibfield
  {journal} {\bibinfo  {journal} {Physical review letters}\ }\textbf {\bibinfo
  {volume} {116}},\ \bibinfo {pages} {090402} (\bibinfo {year}
  {2016})}\BibitemShut {NoStop}%
\bibitem [{\citenamefont {Scala}\ \emph {et~al.}(2013)\citenamefont {Scala},
  \citenamefont {Kim}, \citenamefont {Morley}, \citenamefont {Barker},\ and\
  \citenamefont {Bose}}]{scala2013matter}%
  \BibitemOpen
  \bibfield  {author} {\bibinfo {author} {\bibfnamefont {M.}~\bibnamefont
  {Scala}}, \bibinfo {author} {\bibfnamefont {M.}~\bibnamefont {Kim}}, \bibinfo
  {author} {\bibfnamefont {G.}~\bibnamefont {Morley}}, \bibinfo {author}
  {\bibfnamefont {P.}~\bibnamefont {Barker}}, \ and\ \bibinfo {author}
  {\bibfnamefont {S.}~\bibnamefont {Bose}},\ }\href@noop {} {\bibfield
  {journal} {\bibinfo  {journal} {Physical review letters}\ }\textbf {\bibinfo
  {volume} {111}},\ \bibinfo {pages} {180403} (\bibinfo {year}
  {2013})}\BibitemShut {NoStop}%
\bibitem [{\citenamefont {Romero-Isart}(2017)}]{romero2017coherent}%
  \BibitemOpen
  \bibfield  {author} {\bibinfo {author} {\bibfnamefont {O.}~\bibnamefont
  {Romero-Isart}},\ }\href@noop {} {\bibfield  {journal} {\bibinfo  {journal}
  {New Journal of Physics}\ }\textbf {\bibinfo {volume} {19}},\ \bibinfo
  {pages} {123029} (\bibinfo {year} {2017})}\BibitemShut {NoStop}%
\bibitem [{\citenamefont {Pino}\ \emph {et~al.}(2018)\citenamefont {Pino},
  \citenamefont {Prat-Camps}, \citenamefont {Sinha}, \citenamefont
  {Venkatesh},\ and\ \citenamefont {Romero-Isart}}]{pino2018chip}%
  \BibitemOpen
  \bibfield  {author} {\bibinfo {author} {\bibfnamefont {H.}~\bibnamefont
  {Pino}}, \bibinfo {author} {\bibfnamefont {J.}~\bibnamefont {Prat-Camps}},
  \bibinfo {author} {\bibfnamefont {K.}~\bibnamefont {Sinha}}, \bibinfo
  {author} {\bibfnamefont {B.~P.}\ \bibnamefont {Venkatesh}}, \ and\ \bibinfo
  {author} {\bibfnamefont {O.}~\bibnamefont {Romero-Isart}},\ }\href@noop {}
  {\bibfield  {journal} {\bibinfo  {journal} {Quantum Science and Technology}\
  }\textbf {\bibinfo {volume} {3}},\ \bibinfo {pages} {025001} (\bibinfo {year}
  {2018})}\BibitemShut {NoStop}%
\bibitem [{\citenamefont {Bose}\ \emph {et~al.}(2017)\citenamefont {Bose},
  \citenamefont {Mazumdar}, \citenamefont {Morley}, \citenamefont {Ulbricht},
  \citenamefont {Toro{\v{s}}}, \citenamefont {Paternostro}, \citenamefont
  {Geraci}, \citenamefont {Barker}, \citenamefont {Kim},\ and\ \citenamefont
  {Milburn}}]{bose2017spin}%
  \BibitemOpen
  \bibfield  {author} {\bibinfo {author} {\bibfnamefont {S.}~\bibnamefont
  {Bose}}, \bibinfo {author} {\bibfnamefont {A.}~\bibnamefont {Mazumdar}},
  \bibinfo {author} {\bibfnamefont {G.~W.}\ \bibnamefont {Morley}}, \bibinfo
  {author} {\bibfnamefont {H.}~\bibnamefont {Ulbricht}}, \bibinfo {author}
  {\bibfnamefont {M.}~\bibnamefont {Toro{\v{s}}}}, \bibinfo {author}
  {\bibfnamefont {M.}~\bibnamefont {Paternostro}}, \bibinfo {author}
  {\bibfnamefont {A.~A.}\ \bibnamefont {Geraci}}, \bibinfo {author}
  {\bibfnamefont {P.~F.}\ \bibnamefont {Barker}}, \bibinfo {author}
  {\bibfnamefont {M.}~\bibnamefont {Kim}}, \ and\ \bibinfo {author}
  {\bibfnamefont {G.}~\bibnamefont {Milburn}},\ }\href@noop {} {\bibfield
  {journal} {\bibinfo  {journal} {Physical review letters}\ }\textbf {\bibinfo
  {volume} {119}},\ \bibinfo {pages} {240401} (\bibinfo {year}
  {2017})}\BibitemShut {NoStop}%
\bibitem [{\citenamefont {Schrinski}\ \emph {et~al.}(2020)\citenamefont
  {Schrinski}, \citenamefont {Hornberger},\ and\ \citenamefont
  {Nimmrichter}}]{schrinski2020rule}%
  \BibitemOpen
  \bibfield  {author} {\bibinfo {author} {\bibfnamefont {B.}~\bibnamefont
  {Schrinski}}, \bibinfo {author} {\bibfnamefont {K.}~\bibnamefont
  {Hornberger}}, \ and\ \bibinfo {author} {\bibfnamefont {S.}~\bibnamefont
  {Nimmrichter}},\ }\href@noop {} {\bibfield  {journal} {\bibinfo  {journal}
  {arXiv preprint arXiv:2008.13580}\ } (\bibinfo {year} {2020})}\BibitemShut
  {NoStop}%
\bibitem [{Note1()}]{Note1}%
  \BibitemOpen
  \bibinfo {note} {The state of the particle when it is released from the
  optical trap can be model, to a good approximation, by a thermal harmonic
  oscillator state (see also Supplemental in~\cite {bateman2014near}). Assuming
  $\sigma _{x}$ to be the spatial spread of the state just after the particle
  is release from the optical trap, after a time $\tau $ the spread, due to the
  free evolution, will be $ \sigma _{x}(\tau )= \protect \sqrt {\sigma
  _{x}+\protect \frac {\hbar ^{2} \tau ^2 }{4 m^{2} \sigma _{x}} }. $ Knowing
  that to cover n slits of width d, we need $\sigma _{x}(\tau )\ge n d$, and
  recalling that in our set-up $\sigma _{x}\ll d$ we end up with $ \tau \ge
  \protect \frac {2 n d}{\hbar m \sigma _{x}} $ or equivalently $ \tau \ge
  \protect \frac {4\pi n t_{T}}{d \sigma _{x}}, $ where $t_{T}$ represents the
  Talbot time.}\BibitemShut {Stop}%
\bibitem [{\citenamefont {Xu}\ \emph {et~al.}(2019)\citenamefont {Xu},
  \citenamefont {Jaffe}, \citenamefont {Panda}, \citenamefont {Kristensen},
  \citenamefont {Clark},\ and\ \citenamefont {M{\"u}ller}}]{xu2019probing}%
  \BibitemOpen
  \bibfield  {author} {\bibinfo {author} {\bibfnamefont {V.}~\bibnamefont
  {Xu}}, \bibinfo {author} {\bibfnamefont {M.}~\bibnamefont {Jaffe}}, \bibinfo
  {author} {\bibfnamefont {C.~D.}\ \bibnamefont {Panda}}, \bibinfo {author}
  {\bibfnamefont {S.~L.}\ \bibnamefont {Kristensen}}, \bibinfo {author}
  {\bibfnamefont {L.~W.}\ \bibnamefont {Clark}}, \ and\ \bibinfo {author}
  {\bibfnamefont {H.}~\bibnamefont {M{\"u}ller}},\ }\href@noop {} {\bibfield
  {journal} {\bibinfo  {journal} {Science}\ }\textbf {\bibinfo {volume}
  {366}},\ \bibinfo {pages} {745} (\bibinfo {year} {2019})}\BibitemShut
  {NoStop}%
\bibitem [{\citenamefont {Gabrielse}\ \emph {et~al.}(1990)\citenamefont
  {Gabrielse}, \citenamefont {Fei}, \citenamefont {Orozco}, \citenamefont
  {Tjoelker}, \citenamefont {Haas}, \citenamefont {Kalinowsky}, \citenamefont
  {Trainor},\ and\ \citenamefont {Kells}}]{gabrielse1990thousandfold}%
  \BibitemOpen
  \bibfield  {author} {\bibinfo {author} {\bibfnamefont {G.}~\bibnamefont
  {Gabrielse}}, \bibinfo {author} {\bibfnamefont {X.}~\bibnamefont {Fei}},
  \bibinfo {author} {\bibfnamefont {L.}~\bibnamefont {Orozco}}, \bibinfo
  {author} {\bibfnamefont {R.}~\bibnamefont {Tjoelker}}, \bibinfo {author}
  {\bibfnamefont {J.}~\bibnamefont {Haas}}, \bibinfo {author} {\bibfnamefont
  {H.}~\bibnamefont {Kalinowsky}}, \bibinfo {author} {\bibfnamefont
  {T.}~\bibnamefont {Trainor}}, \ and\ \bibinfo {author} {\bibfnamefont
  {W.}~\bibnamefont {Kells}},\ }\href@noop {} {\bibfield  {journal} {\bibinfo
  {journal} {Physical review letters}\ }\textbf {\bibinfo {volume} {65}},\
  \bibinfo {pages} {1317} (\bibinfo {year} {1990})}\BibitemShut {NoStop}%
\bibitem [{\citenamefont {Schwarz}\ \emph {et~al.}(2012)\citenamefont
  {Schwarz}, \citenamefont {Versolato}, \citenamefont {Windberger},
  \citenamefont {Brunner}, \citenamefont {Ballance}, \citenamefont {Eberle},
  \citenamefont {Ullrich}, \citenamefont {Schmidt}, \citenamefont {Hansen},
  \citenamefont {Gingell} \emph {et~al.}}]{schwarz2012cryogenic}%
  \BibitemOpen
  \bibfield  {author} {\bibinfo {author} {\bibfnamefont {M.}~\bibnamefont
  {Schwarz}}, \bibinfo {author} {\bibfnamefont {O.}~\bibnamefont {Versolato}},
  \bibinfo {author} {\bibfnamefont {A.}~\bibnamefont {Windberger}}, \bibinfo
  {author} {\bibfnamefont {F.}~\bibnamefont {Brunner}}, \bibinfo {author}
  {\bibfnamefont {T.}~\bibnamefont {Ballance}}, \bibinfo {author}
  {\bibfnamefont {S.}~\bibnamefont {Eberle}}, \bibinfo {author} {\bibfnamefont
  {J.}~\bibnamefont {Ullrich}}, \bibinfo {author} {\bibfnamefont {P.~O.}\
  \bibnamefont {Schmidt}}, \bibinfo {author} {\bibfnamefont {A.~K.}\
  \bibnamefont {Hansen}}, \bibinfo {author} {\bibfnamefont {A.~D.}\
  \bibnamefont {Gingell}},  \emph {et~al.},\ }\href@noop {} {\bibfield
  {journal} {\bibinfo  {journal} {Review of Scientific Instruments}\ }\textbf
  {\bibinfo {volume} {83}},\ \bibinfo {pages} {083115} (\bibinfo {year}
  {2012})}\BibitemShut {NoStop}%
\bibitem [{Note2()}]{Note2}%
  \BibitemOpen
  \bibinfo {note} {{ Note that a Si/SiO2 particle at cryogenic temperature, can
  easily reach internal temperature above 200~K if exposed for 1~s to light at
  wavelength of 1550~nm focused with a $0.8 NA$ lens with intensity $I =
  90$~mW/$\mu $m and reach internal temperature of more than 500~K if at room
  temperature.}}\BibitemShut {Stop}%
\bibitem [{Note3()}]{Note3}%
  \BibitemOpen
  \bibinfo {note} {We have estimated a reduction of the visibility less than
  $0.0002 \%$ due to the presence of thermal photon emission decoherence for
  both Si and SiO2 particles with mass of $10^7$~amu in the experimental setup
  describe by the parameters in Tab.~\ref {tab:parameters} with the cryogenic
  environmental temperature of $4$~K.}\BibitemShut {Stop}%
\bibitem [{\citenamefont {Ralph}\ \emph {et~al.}(2018)\citenamefont {Ralph},
  \citenamefont {Toro{\v{s}}}, \citenamefont {Maskell}, \citenamefont {Jacobs},
  \citenamefont {Rashid}, \citenamefont {Setter},\ and\ \citenamefont
  {Ulbricht}}]{ralph2018dynamical}%
  \BibitemOpen
  \bibfield  {author} {\bibinfo {author} {\bibfnamefont {J.~F.}\ \bibnamefont
  {Ralph}}, \bibinfo {author} {\bibfnamefont {M.}~\bibnamefont {Toro{\v{s}}}},
  \bibinfo {author} {\bibfnamefont {S.}~\bibnamefont {Maskell}}, \bibinfo
  {author} {\bibfnamefont {K.}~\bibnamefont {Jacobs}}, \bibinfo {author}
  {\bibfnamefont {M.}~\bibnamefont {Rashid}}, \bibinfo {author} {\bibfnamefont
  {A.~J.}\ \bibnamefont {Setter}}, \ and\ \bibinfo {author} {\bibfnamefont
  {H.}~\bibnamefont {Ulbricht}},\ }\href@noop {} {\bibfield  {journal}
  {\bibinfo  {journal} {Physical Review A}\ }\textbf {\bibinfo {volume} {98}},\
  \bibinfo {pages} {010102} (\bibinfo {year} {2018})}\BibitemShut {NoStop}%
\bibitem [{\citenamefont {Schrinski}\ \emph {et~al.}(2019)\citenamefont
  {Schrinski}, \citenamefont {Nimmrichter}, \citenamefont {Stickler},\ and\
  \citenamefont {Hornberger}}]{schrinski2019macroscopicity}%
  \BibitemOpen
  \bibfield  {author} {\bibinfo {author} {\bibfnamefont {B.}~\bibnamefont
  {Schrinski}}, \bibinfo {author} {\bibfnamefont {S.}~\bibnamefont
  {Nimmrichter}}, \bibinfo {author} {\bibfnamefont {B.~A.}\ \bibnamefont
  {Stickler}}, \ and\ \bibinfo {author} {\bibfnamefont {K.}~\bibnamefont
  {Hornberger}},\ }\href@noop {} {\bibfield  {journal} {\bibinfo  {journal}
  {Physical Review A}\ }\textbf {\bibinfo {volume} {100}},\ \bibinfo {pages}
  {032111} (\bibinfo {year} {2019})}\BibitemShut {NoStop}%
\bibitem [{\citenamefont {Hebestreit}\ \emph {et~al.}(2018)\citenamefont
  {Hebestreit}, \citenamefont {Frimmer}, \citenamefont {Reimann},\ and\
  \citenamefont {Novotny}}]{hebestreit2018sensing}%
  \BibitemOpen
  \bibfield  {author} {\bibinfo {author} {\bibfnamefont {E.}~\bibnamefont
  {Hebestreit}}, \bibinfo {author} {\bibfnamefont {M.}~\bibnamefont {Frimmer}},
  \bibinfo {author} {\bibfnamefont {R.}~\bibnamefont {Reimann}}, \ and\
  \bibinfo {author} {\bibfnamefont {L.}~\bibnamefont {Novotny}},\ }\href@noop
  {} {\bibfield  {journal} {\bibinfo  {journal} {Physical review letters}\
  }\textbf {\bibinfo {volume} {121}},\ \bibinfo {pages} {063602} (\bibinfo
  {year} {2018})}\BibitemShut {NoStop}%
\bibitem [{\citenamefont {Bongs}\ \emph {et~al.}(2019)\citenamefont {Bongs},
  \citenamefont {Holynski}, \citenamefont {Vovrosh}, \citenamefont {Bouyer},
  \citenamefont {Condon}, \citenamefont {Rasel}, \citenamefont {Schubert},
  \citenamefont {Schleich},\ and\ \citenamefont {Roura}}]{bongs2019taking}%
  \BibitemOpen
  \bibfield  {author} {\bibinfo {author} {\bibfnamefont {K.}~\bibnamefont
  {Bongs}}, \bibinfo {author} {\bibfnamefont {M.}~\bibnamefont {Holynski}},
  \bibinfo {author} {\bibfnamefont {J.}~\bibnamefont {Vovrosh}}, \bibinfo
  {author} {\bibfnamefont {P.}~\bibnamefont {Bouyer}}, \bibinfo {author}
  {\bibfnamefont {G.}~\bibnamefont {Condon}}, \bibinfo {author} {\bibfnamefont
  {E.}~\bibnamefont {Rasel}}, \bibinfo {author} {\bibfnamefont
  {C.}~\bibnamefont {Schubert}}, \bibinfo {author} {\bibfnamefont {W.~P.}\
  \bibnamefont {Schleich}}, \ and\ \bibinfo {author} {\bibfnamefont
  {A.}~\bibnamefont {Roura}},\ }\href@noop {} {\bibfield  {journal} {\bibinfo
  {journal} {Nature Reviews Physics}\ ,\ \bibinfo {pages} {1}} (\bibinfo {year}
  {2019})}\BibitemShut {NoStop}%
\bibitem [{\citenamefont {Hempston}\ \emph {et~al.}(2017)\citenamefont
  {Hempston}, \citenamefont {Vovrosh}, \citenamefont {Toro{\v{s}}},
  \citenamefont {Winstone}, \citenamefont {Rashid},\ and\ \citenamefont
  {Ulbricht}}]{hempston2017force}%
  \BibitemOpen
  \bibfield  {author} {\bibinfo {author} {\bibfnamefont {D.}~\bibnamefont
  {Hempston}}, \bibinfo {author} {\bibfnamefont {J.}~\bibnamefont {Vovrosh}},
  \bibinfo {author} {\bibfnamefont {M.}~\bibnamefont {Toro{\v{s}}}}, \bibinfo
  {author} {\bibfnamefont {G.}~\bibnamefont {Winstone}}, \bibinfo {author}
  {\bibfnamefont {M.}~\bibnamefont {Rashid}}, \ and\ \bibinfo {author}
  {\bibfnamefont {H.}~\bibnamefont {Ulbricht}},\ }\href@noop {} {\bibfield
  {journal} {\bibinfo  {journal} {Applied Physics Letters}\ }\textbf {\bibinfo
  {volume} {111}},\ \bibinfo {pages} {133111} (\bibinfo {year}
  {2017})}\BibitemShut {NoStop}%
\bibitem [{\citenamefont {Frimmer}\ \emph {et~al.}(2017)\citenamefont
  {Frimmer}, \citenamefont {Luszcz}, \citenamefont {Ferreiro}, \citenamefont
  {Jain}, \citenamefont {Hebestreit},\ and\ \citenamefont
  {Novotny}}]{frimmer2017controlling}%
  \BibitemOpen
  \bibfield  {author} {\bibinfo {author} {\bibfnamefont {M.}~\bibnamefont
  {Frimmer}}, \bibinfo {author} {\bibfnamefont {K.}~\bibnamefont {Luszcz}},
  \bibinfo {author} {\bibfnamefont {S.}~\bibnamefont {Ferreiro}}, \bibinfo
  {author} {\bibfnamefont {V.}~\bibnamefont {Jain}}, \bibinfo {author}
  {\bibfnamefont {E.}~\bibnamefont {Hebestreit}}, \ and\ \bibinfo {author}
  {\bibfnamefont {L.}~\bibnamefont {Novotny}},\ }\href@noop {} {\bibfield
  {journal} {\bibinfo  {journal} {Physical Review A}\ }\textbf {\bibinfo
  {volume} {95}},\ \bibinfo {pages} {061801} (\bibinfo {year}
  {2017})}\BibitemShut {NoStop}%
\bibitem [{\citenamefont {Hebestreit}(2017)}]{hebestreit2017thermal}%
  \BibitemOpen
  \bibfield  {author} {\bibinfo {author} {\bibfnamefont {E.}~\bibnamefont
  {Hebestreit}},\ }\emph {\bibinfo {title} {Thermal properties of levitated
  nanoparticles}},\ \href@noop {} {Ph.D. thesis},\ \bibinfo  {school} {ETH
  Zurich} (\bibinfo {year} {2017})\BibitemShut {NoStop}%
\end{thebibliography}%

\pagebreak
\widetext
\begin{center}
\textbf{\large Supplemental Materials:Prospects for near-field interferometric tests of Collapse Models}
\end{center}
\setcounter{equation}{0}
\setcounter{section}{0}
\setcounter{figure}{0}
\setcounter{table}{0}
\setcounter{page}{1}
\makeatletter
\renewcommand{\theequation}{S\arabic{equation}}
\renewcommand{\thefigure}{S\arabic{figure}}

In this supplemental material we give additional details on the derivation of the interference pattern probability in Eq.~\eqref{pattern} of the main text. In particular, we report the expressions for the generalized Talbot coefficients ($B_n$), the free-fall environemnetal decoherence kernels $R_n$, and the kernels describing the action of the CSL fundamental noise during the free evolution of the nanoparticles. For additional details, we refer the reader to Refs.~\cite{bateman2014near,belenchia2019talbot}.

\section{Generalized Talbot Coefficients}
Here we report the expression for the generalized Talbot coefficients used in our simulations. We do not go in to the details of their derivation and refer to~\cite{belenchia2019talbot} and reference therein for an exhaustive analysis of the Talbot effect. 
The generalized Talbot coefficients characterizing the interference pattern produced by the interaction of a dielectric spherical particle with a laser grating are given by

\begin{align}
B_{n}\left(\frac{s}{d}\right)= e^{F-c_{\rm{abs}}/2} \sum_{k=-\infty}^{\infty}\left(\frac{\zeta_{\rm{coh}}+a+c_{\rm{abs}}/2}{\zeta_{\rm{coh}}-a-c_{\rm{abs}}/2}\right)^{\frac{n+k}{2}}J_{k}(b)J_{n+k}\left(\sign(\zeta_{\rm{coh}}-a-c_{\rm{abs}}/2)\sqrt{\zeta_{\rm{coh}}^{2}-(a+c_{\rm{abs}}/2)^{2}}\right).
\end{align}
with    
\begin{align}\label{abF}
c_{\rm{abs}}&=
\frac{4\sigma_{abs}}{ h c}\frac{E_{L}}{a_{L}}(1-\cos(\pi s/d)\nonumber\\
\zeta_{coh} &= \frac{16F_{z}(-\lambda/8) E_L}{\hbar a_L k I_{0}}\sin\left(\frac{\pi s}{d}\right)\nonumber\\
a(s) &= \int d\tau \frac{8\pi  E_{L}}{\hbar \omega a_{L}}\int d \Omega\,{\rm Re}\big(f^{*}(k , k \mathbf{n})f(- k , k\mathbf{n})\big)[\cos(k n_{z} s)-\cos(ks)],\nonumber\\
b(s)&=  \int d\tau\frac{8\pi  E_{L}}{\hbar  \omega a_{L}}\int d \Omega \,{\rm Im} \big(f^{*}(k , k \mathbf{n})f(- k , k\mathbf{n})\big)\sin(kn_{z}s),\nonumber\\
F(s)&=  \int d\tau \frac{8 \pi E_{L}}{\hbar  \omega a_{L}}\int d \Omega\, |f(k,k\mathbf{n})|^{2} [\cos((1-n_{z})ks)-1].
\end{align}
where $\lambda= 2\pi / k= 2\pi /\omega c $ is the light wavelength, $E_{L}$ is the laser pulse's energy, $I_{0}$ and $a_{L}$ are respectively the intensity parameter and the spot area of the laser, $F_{z}(z)$ is the longitudinal light-induced force on the dielectric sphere, $\sigma_{abs}$ is the photon absorption cross section, and $f(k,k\mathbf{n})$ the photon scattering cross section.

\section{Environmental Decoherence}
As discussed in the main text, our simulations account for several sources of environmental decoherence acting during the free fall times in the experiment.  In particular, we account for decoherence due to collision with the residual gas in the vacuum chamber; black-body radiation, emission, scattering, and absorption -- taking into account the heating of the particle (photonic environment) during the trapping period and the subsequent cooling during free-fall. Here we report the expression for the kernel $R_{n}$ in Eq.~(3) in the main text that account for all these decoherence sources.
\begin{align}
\ln(R_{n})=& -\Gamma_{\text{coll}}(t_{1}+t_{2})+\int d\omega \gamma_{\text{abs}}(\omega)\left[\frac{\text{Si}(a_{n})}{a_{n}}-1\right](t_{1}+t_{2})+\int d\omega \gamma_{\text{sca}}(\omega)\left[\frac{\text{Si}(2a_{n})}{a_{n}}-\text{sinc}^{2}(a_{n})-1\right](t_{1}-t_{2})\nonumber\\
&\int d\omega \int_{0}^{1} d\theta \{t_{1}\gamma_{\text{emi}}[\omega,T_{\text{int}}(t_{1}-t_{1}\theta)]+t_{2}\gamma_{\text{emi}}[\omega,T_{\text{int}(t_{1}+t_{2}\theta)] }[\text{sinc}(a_{n}\theta)-1]
\end{align}
with $a_{n}= n\,h\,\omega\, t_{2} t_{1} / ((t_2+t_1)m\,c\,d) $, and $\text{Si}$ the sine integral. The collision rate $\Gamma_\text{coll}$ is given by
\begin{align}
\Gamma_{\text{coll}} = \frac{ 4\pi \Gamma(9/10)}{5\sin(\pi/5)}\left(\frac{3\pi C_{6}}{2\hbar}^{2/5} \frac{p_{g}v_{g}}{k_{\text{B}}T_{\text{env}}}\right)
\end{align}
while the scattering, emission and absorption rates by
\begin{align}
\gamma_{\text{sca/abs}}(\omega) =\frac{ (\omega / \pi c)^{2}\sigma_{\text{sca/abs}}(\omega)}{\exp(\hbar \omega/k_{\text{B}}T_{\text{env}})-1},\hspace{1cm}
&\gamma_{\text{emi}}[\omega, T_{\text{int}}]= \left(\frac{\omega}{\pi c}\right)^{2}\sigma_{\text{abs}}\exp\left(-\frac{\hbar \omega}{k_{\text{B}}T_{\text{int}}}\right)\text{Im}\left\{\frac{\varepsilon(\omega)-1}{\varepsilon(\omega)+2}\right\}
\end{align}
where  $v_{g}$ and $p_{g}$ are the mean velocity and pressure of the gas, $T_{\text{env}}$  the environmental temperature, $\sigma_{\text{abs/sca}}$ the photon scattering/absorption cross section, $\varepsilon(\omega)$ the electric permittivity, and 
\begin{align}
C_{6} \simeq \frac{ 3 \alpha(\omega=0) \alpha_{g} I_{g}I}{32 \pi^{2}\varepsilon_{0}^{2}(I_{g}+I)}
\end{align}
the van der Waals coupling constant where $\alpha$, $\alpha_{g}$ are the static polarizabilities and $I$, $I_g$ the ionization energies of the nanosphere and the gas particle, respectively.

We refer the reader to the supplemental information of~\cite{bateman2014near} and the reference therein for a detailed derivation of these expressions.

\section{ CSL-Decoherence}
The dissipative term describing the effective decoherence of the center-of-mass wavefunction -- for the reduced one-dimensional state of motion of a nanoparticle -- due to the CSL reads
\begin{align}\label{cslme}
\mathcal{L}_{CSL}(x,x')=-\frac{\lambda_{CSL}(4\pi r_C^2)^{3/2}}{(2\pi\hbar)^{3}}\int d\mathbf{q}\,\frac{\tilde{\mu}(\mathbf{q})^2}{m_0^2}\,e^{-r_C^2\mathbf{q}^{2}/\hbar^{2}}\,\left(e^{-\frac{i}{\hbar}{q}_{z}({x}-{x'})}-1\right)
\end{align}   
in position representation. Here $m_{0}$ the nucleon mass , $\lambda_{CSL}$ and $r_{c}$ the rate and the localization length of the CSL model, and $\mu(\bf{q})$ the Fourier transform of the nano-particle mass density $\mu(\bf{x})$
,i.e.
\begin{align}
\tilde{\mu}(\mathbf{q}):= \int e^{-\frac{i}{\hbar} \mathbf{q}\cdot\mathbf{x}}\mu(\mathbf{x}).
\end{align}
that in the case of a homogeneous and spherical mass distribution of radius $R$ is given by
\begin{align}
 \tilde{\mu}(\mathbf{q})= \frac{4 \pi \hbar R^2}{q} J_{1}(q R/\hbar)
\end{align}
where $J_{1}(q)$ denotes the spherical Bessel function of the first kind.  
Exploiting this equation, rewriting the integral in Eq.~\eqref{cslme} in spherical coordinates, and performing the integration over the solid angle the CSL dissipative term simplifies to 
\begin{align}
\mathcal{L}_{CSL}(\mathbf{x},\mathbf{x}')= \frac{64 \pi^{3/2}\,\lambda_{CSL}\, rc^{3} R^{4}}{\hbar\, m_{0}^{2}}\int_{0}^{\infty} d q\, e^{-r_{c}q^2/\hbar^2} J_{1}(q R/\hbar)^{2}\left(\text{sinc}(q|{x}'-{x}|)/\hbar)-1 \right).
\end{align}
This results in the following, additional kernels entering the expression for the interference pattern probability:
\begin{align}
R_{n}^{\text{CSL}}= \exp\left\{ \Gamma_{\text{CSL}}\left(f_{\text{CSL}}\left(\frac{ h\, n\, q\, t_1 t_2}{m\, d (t_1+t_2)}\right)-1\right)(t_1+t_2)\right\}
\end{align}
where $d$ is again the grating period and 
\begin{align}
\Gamma_{\text{CSL}}&=  \sqrt{\frac{32}{\pi}}\frac{\lambda_{\text{CSL}}\, r_{c}^{3}}{\hbar^{3} m_{0}^{2}} \int dq q^2 e^{-r_{c}^{2}q^2/\hbar^2}\tilde{\rho}(q)^{2}  \nonumber\\
f_{\text{CSL}}(x)&=\sqrt{\frac{32 }{\pi}} \frac{\lambda_{CSL}\,r_{c}^{3}}{\hbar^3\,\Gamma_{\text{CSL}}}\int dq q^{2} e^{-r_{c}^{2}q^{2}/\hbar^{2}} \tilde{\rho}(q)^{2} \text{Si}\left(\frac{q x}{\hbar}\right) =\frac{64 \pi^{3/2}\lambda_{CSL} rc^{3} R^{4} }{\hbar\,\Gamma_{\text{CSL}}} \int_{0}^{\infty}dq e^{-q^2 r_{c}^{2}/\hbar^{2}}J_{1}\left(\frac{q R}{\hbar}\right)^{2}\text{Si}\left(\frac{qx}{\hbar}\right)^{2}.
\end{align}

\end{document}